\documentclass[twocolumn]{aa}
\usepackage{graphicx}
\usepackage{epsfig,graphicx,rotating,fancyhdr,longtable}
\usepackage{multirow}
\usepackage{txfonts}

\usepackage{stfloats}
\begin{document}

\title{GRB171205A/SN2017iuk: A local low-luminosity gamma-ray burst}
\author{V. D'Elia$^1$, S. Campana$^{2}$, A. D'A\`i$^{3}$, M. De Pasquale$^{4}$, S. W. K. Emery$^{5}$, D. D. Frederiks$^6$, A. Lien$^{7,8}$, A. Melandri$^{2}$, K. L. Page$^{9}$, R. L. C. Starling$^{9}$, D. N. Burrows$^{10}$, A. A. Breeveld$^{5}$, S. R. Oates$^{11}$, P. T. O'Brien$^{9}$, J. P. Osborne$^{9}$, M. H. Siegel$^{10}$, G. Tagliaferri$^{2}$, P. J. Brown$^{12}$, S. B. Cenko$^{13,14}$, D. S. Svinkin$^6$, A. Tohuvavohu$^{10}$, A. E. Tsvetkova$^6$}

\institute
{$^1$Space Science Data Center - Agenzia Spaziale Italiana, via del Politecnico, s.n.c., I-00133, Roma, Italy\\
$^{2}$INAF-Osservatorio Astronomico di Brera, via Bianchi 46, I-23807, Merate, Italy\\
$^3$INAF - IASF Palermo, via Ugo La Malfa, 153, 90146 - Palermo, Italy\\
$^{4}$Department of Astronomy and Space Sciences, Istanbul University, Beyazit, 34119, Istanbul, Turkey\\
$^5$Mullard Space Science Laboratory, University College London, Holmbury St Mary, Dorking, Surrey RH5 6NT, UK\\
$^6$Ioffe Physical-Technical Institute, Politekhnicheskaya 26, St. Petersburg 194021, Russia\\
$^7$Center for Research and Exploration in Space Science and Technology (CRESST) and
NASA Goddard Space Flight Center, Greenbelt, MD 20771, USA\\
$^8$Department of Physics, University of Maryland, Baltimore County, 1000 Hilltop Circle, Baltimore, MD 21250, USA\\
$^9$Department of Physics and Astronomy, University of Leicester, University Road, Leicester LE7 1RH, UK\\
$^{10}$Department of Astronomy \& Astrophysics, The Pennsylvania State University, University Park, PA 16802, USA\\
$^{11}$Department of Physics, University of Warwick, Coventry, CV4 7AL, UK\\
$^{12}$George P. and Cynthia Woods Mitchell Institute for Fundamental Physics \& Astronomy, Texas A. \& M. University, Department of Physics and Astronomy, 4242 TAMU, College Station, TX 77843, USA\\
$^{13}$Astrophysics Science Division, NASA Goddard Space Flight Center, Mail Code 661, Greenbelt, MD 20771, USA\\
$^{14}$Joint Space-Science Institute, University of Maryland, College Park, MD 20742, USA\\
}

 \abstract 
  {Gamma-ray bursts (GRBs) occurring in the local Universe constitute an interesting sub-class of the GRB family, since their luminosity is on average lower than that of their cosmological analogs. Attempts to understand in a global way this peculiar behaviour is still not possible, since the sample of low redshift GRBs is small, and the properties of individual objects are too different from each other. In addition, their closeness (and consequently high fluxes) make these sources ideal targets for extensive follow-up even with small telescopes, considering also that these GRBs are conclusively associated with supernova (SN) explosions.} 
  {We aim to contribute to the study of local bursts by reporting the case of GRB 171205A. This source was discovered by {\it Swift}\, Burst Alert Telescope (BAT) on 2017, December 5 and soon associated with a low redshift host galaxy ($z=0.037$), and an emerging SN (SN 2017iuk).}
  {We analyzed the full {\it Swift}\, dataset, comprising the UV-Optical Telescope (UVOT), X-ray Telescope (XRT) and BAT data. In addition, we employed the Konus-Wind high energy data as a valuable extension at $\gamma$-ray energies.}
  {The photometric SN signature is clearly visible in the UVOT $u$, $b$ and $v$ filters. The maximum emission is reached at $\sim 13$ (rest frame) days, and the whole bump resembles that of SN 2006aj, but lower in magnitude and with a shift in time of $+2$ d. A prebump in the $v$-band is also clearly visible, and this is the first time that such a feature is not observed achromatically in GRB-SNe. Its physical origin cannot be easily explained.
  
  The X-ray spectrum shows an intrinsic Hydrogen column density $N_{H,int} = 7.4^{+4.1}_{-3.6}\times 10^{20}$ cm$^{-2}$, which is at the low end of the $N_{H,int}$, even considering just low redshift GRBs. The spectrum also features a thermal component, which is quite common in GRBs associated with SNe, but whose origin is still a matter of debate.
  
  Finally, the isotropic energy in the $\gamma$-ray band, $E_{\rm iso} = 2.18^{+0.63}_{-0.50} \times 10^{49}$ erg, is lower than those of cosmological GRBs. Combining this value with the peak energy in the same band, E$_p=125^{+141}_{-37}$ keV, implies that GRB 171205A is an outlier of the Amati relation, as are some other low redshift GRBs, and its emission mechanism should be different from that of canonical, farther away GRBs.  }
{}

\authorrunning {D'Elia et al.}
\titlerunning {GRB 171205A: a low luminosity GRB}        

\maketitle

\section{Introduction}

Long gamma-ray bursts (LGRBs) are the most powerful stellar explosions since the formation of the Universe. Most of their observed electromagnetic energy is released in the $0.01-1$ MeV band, and they occur at a rate of one to two per day over the whole sky. Their subsequent fading emission, which is observed at multiple wavelengths, is long-lived and called the afterglow (see, e.g., M\'esz\'aros 2006 for a review), can outshine every other known object in the Universe, both in the optical/near-infrared (NIR) (e.g., Racusin et al. 2008) and in the X-rays (Cusumano et al. 2006). Contrary to their short-lived brothers, which last less than two seconds, LGRBs are thought to be the result of the core-collapse of a massive star, under the mechanism named the collapsar model (Woosley 1993; Paczynski 1998; MacFadyen \& Woosley 1999). Alternatively, an exploding massive star in a binary system may cause the companion neutron star to reach the critical mass and to collapse into a black hole launching a GRB (Fryer, Rueda \& Ruffini 2014). Given their extraordinary power output at all wavelengths, LGRBs are detected up to very high redshifts (Tanvir et al. 2009; Salvaterra et al. 2009; Cucchiara et al. 2009), making them powerful probes of the early Universe (see, e.g., Vreeswijk et al. 2001; Berger et al. 2006; Fynbo et al. 2006a; Prochaska et al. 2007; Sparre et al. 2013).

Considering the other edge of the Universe, close-by LGRBs are equally important, because these sources show on average lower luminosities than cosmological events (e.g., GRB 980425 at $z=0.0085$: Galama et al. 1998, Kulkarni et al. 1998, Pian et al. 2006; GRB 031203 at $z=0.105$: Malesani et al. 2004, Soderberg et al. 2004, Watson et al. 2004; GRB 060218 at $z=0.0331$: Campana et al. 2006, Mazzali et al. 2006, Virgili et al. 2009; GRB 100316D at $z=0.059$: Fan et al. 2011, Starling et al. 2011, Bufano et al. 2011), despite some noteworthy exceptions like GRB 130427A (Maselli et al. 2013). Both theoretical (e.g., Daigne \& Mochkovitch 2007; Barniol-Duran et al. 2015) and phenomenological (e.g., Virgili et al. 2008; Dereli et al. 2017) approaches have been attempted to explain the peculiarities of these sources in the framework of the standard GRB model. However, the sample is still sparse and the individual properties of these bursts are too different from each other to draw firm conclusions. 

On the other hand, the connection between LGRBs and supernovae (SNe) is firmly established. This relationship can be directly probed only for low redshift events, as the SN emission becomes too faint to be detected at cosmological distances. In particular, for close-by events ($z<0.3$), a detailed spectroscopic monitoring of the accompanying SN makes it possible to derive physical parameters of the ejecta and the progenitor (Galama et al. 1998; Patat et al. 2001; Hjorth et al. 2003; Stanek et al. 2003; Malesani et al. 2004; Pian et al. 2006; Bufano et al. 2012; Mazzali et al. 2006a,b; Woosley \& Bloom 2006; Hjorth \& Bloom 2012). These observations reveal that SNe accompanying LGRBs are explosions of bare stellar cores, that is, their progenitors (whose estimated mass is higher than $\sim$20 M$_\odot$) have lost all their hydrogen and helium envelopes before collapse (a.k.a. supernovae of type Ic). However, this scenario is challenged by two surprising exceptions: 
%
%
GRB\,060505 at $z=0.089$ and GRB\,060614 at $z=0.125$, which show no evidence for SN emission down to very deep limits, suggesting a new phenomenological type of massive stellar death (Della Valle et al. 2006; Fynbo et al. 2006b).

The GRB field was revolutionized by the {\it Neil Gehrels Swift Observatory} (Gehrels et al. 2004). 
After more than $13$ years of operations, {\it Swift} has detected more than $1000$ GRBs. The key to the success of this mission is its ability to quickly repoint its narrow field instruments and to obtain an arcsecond position of the afterglow in the X-ray and optical/UV bands within a few minutes. 
One of these bursts, GRB 171205A, was detected by {\it Swift} on 2017 December 5 and an association with a close-by galaxy at $z\sim 0.04$ was soon revealed (Izzo et al. 2017a). Given the proximity of this LGRB, the interest of the scientific community was testified by the massive follow-up of the afterglow at all wavelengths (see Section 4.1 for details).

In this work we have concentrated on the {\it Swift} data acquired for this special LGRB. 
The paper is organized as follows: Section 2 summarizes the properties of GRB\,171205A and its association with a supernova, as reported in the literature; Sect. 3 introduces our dataset and illustrates the data reduction process, a subsection is devoted to each of the three {\it Swift} instruments; Sect. 4 presents and discusses our results; in Sect. 5 we draw our conclusions. Unless otherwise stated, we assume a cosmology with $H_0=67.3$ km s$^{-1}$ Mpc$^{-1}$, $\Omega_{\rm m} = 0.315$, $\Omega_\Lambda = 0.685$ (Planck Collaboration 2014).
%
%

\section{GRB\,171205A}

GRB\,171205A was discovered by the Burst Alert Telescope (BAT, Barthelmy et al. 2005)
instrument on board {\it Swift} on 2017 December 5
at 07:20:43.9 UT (D'Elia et al. 2017). The BAT light curve shows some weak emission with multiple overlapping peaks that starts at $T{_0} -40$ s and ends at $T{_0} + 200$ s, where $T_0$ is the burst detection time. The BAT spectrum is best fit by a simple power-law model (Barthelmy et al. 2017). The GRB was also detected by Konus-Wind (Frederiks et al. 2017).
%
%
%
%
The X-Ray Telescope (XRT, Burrows et al. 2005) began observation the GRB  about $150$ s after the trigger. 
(Kennea et al. 2017).  

The Ultraviolet and Optical Telescope (UVOT, Roming et al. 2005) began settled 
 observations of the field of GRB\,171205A $154$ s after the BAT trigger. 
A source consistent with the XRT enhanced position (Osborne et al. 2017)
was detected in the initial UVOT exposures and well detected in all filters.
The revised source position was RA(J2000)=11:09:39.55, Dec.(J2000)=-12:35:17.9 (Emery \& D'Elia 2017).

The afterglow was tentatively associated with a close-by bright spiral host galaxy at $z=0.037$ (2MASX J11093966-1235116) even before its optical position was 
 known (Izzo et al. 2017a). Ground-based facilities pointed to the GRB starting a few minutes
after the BAT notice, allowing for the detection of the afterglow in the optical and near
infrared (e.g., Butler et al. 2017; Mao et al. 2017; Choi et al. 2017; Melandri et al. 2017), and the confirmation of the association with 
 the host galaxy through the detection of both absorption and emission lines at the common redshift of $z=0.0368$ (Izzo et al. 2017b).
The host galaxy mass is $log_{10} M/M_{\sun}= 10.1 \pm 0.1$, which is at least a factor of ten heavier than any other low-redshift GRB with a confirmed supernova (SN) counterpart (Perley \& Taggart 2017), and it has a star-formation rate (SFR) of $3\pm 1$\, M$_{\sun}$/yr.
The detection of spectral features from an emerging SN (SN 2017iuk) was indeed reported two days after the burst (de Ugarte Postigo et al. 2017a), with bumps similar to that seen in the very first stages of SN 1998bw (Patat et al. 2001). 

Finally, a bright afterglow is also detected in several radio bands (de Ugarte Postigo et al. 2017b; Smith \& Tanvir 2017; Perley et al. 2017; 
Trushkin et al. 2017). As reported by de Ugarte Postigo et al. (2017b) this is the second brightest GRB ever detected at these wavelengths.

\section{Data reduction and analysis}

\subsection{$\gamma$-ray data reduction and analysis}
\subsubsection{BAT data reduction and analysis}

\begin{figure}[!h]
\includegraphics[width=\columnwidth]{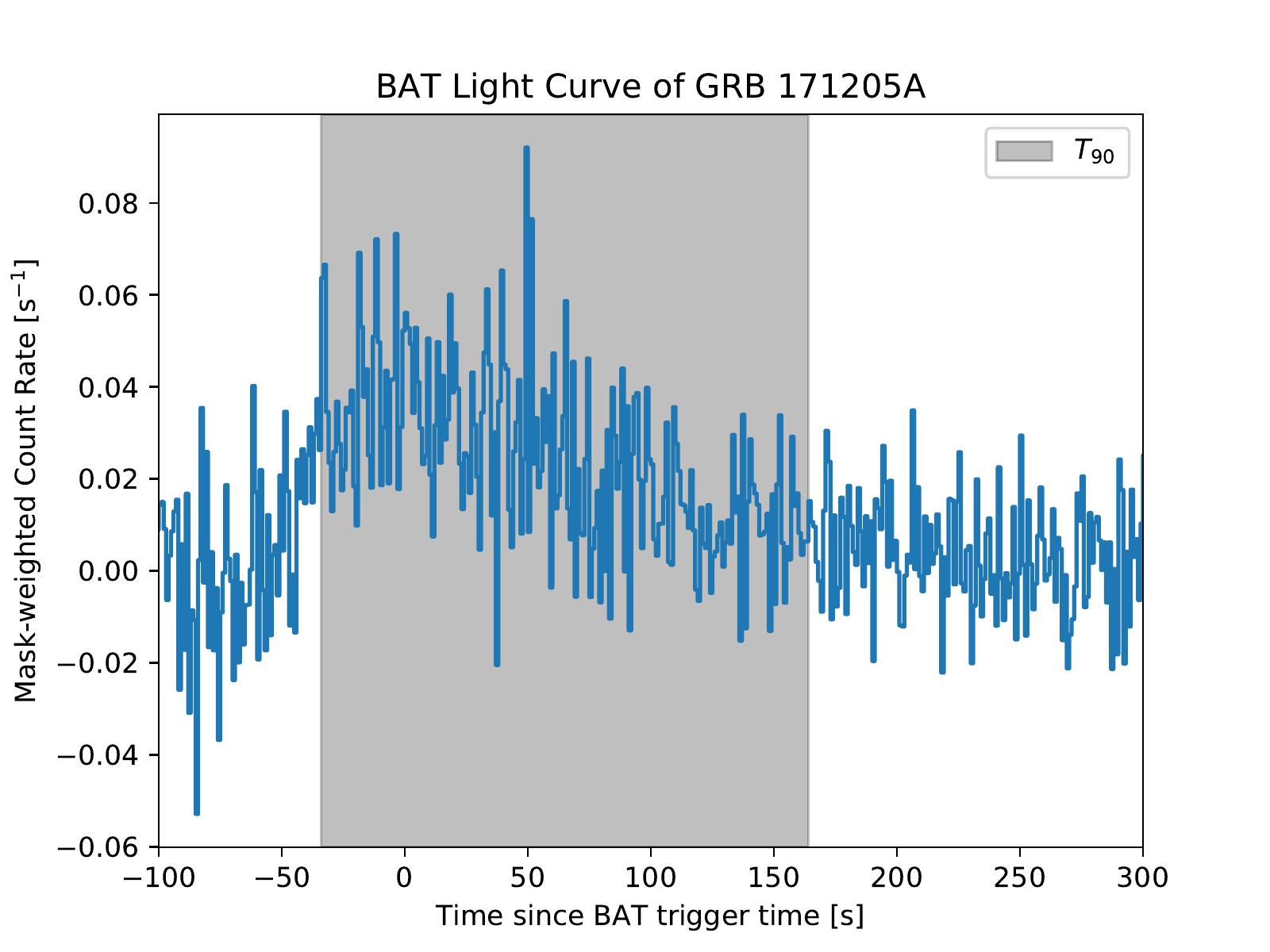}
%
%
%
\caption{{\it Swift}/BAT mask-weighted light curve in 15-350 keV and 1-s time bins. The gray region encloses the $T_{90}$ interval.}
\label{fig:BAT_light_curve}
\end{figure}

The BAT data analysis uses the event-by-event data collected from T$_0-239$ s to T$_0+963$ s, the standard BAT software (HEASOFT 6.22.1\footnote{http://heasarc.nasa.gov/lheasoft/}) and the latest calibration database (CALDB\footnote{http://heasarc.gsfc.nasa.gov/docs/heasarc/caldb/swift/}). The BAT mask-weighted light curve (Fig.~\ref{fig:BAT_light_curve}) shows some weak overlapping pulses that start at $\sim$ T$_0-40$ s and end at $\sim$ T$_0+200$ s. $T_{90}$ (15-350 keV) is 190.5 $\pm$ 33.9 s (estimated error including systematics). In the following, $T_{90}$ ($T_{100}$) indicates the time interval where integrated counts from the GRB raise from $5$\% to $95$\% ($0$\% to $100$\%).
%
%
%
%

\begin{figure}[!h]
\includegraphics[width=\columnwidth]{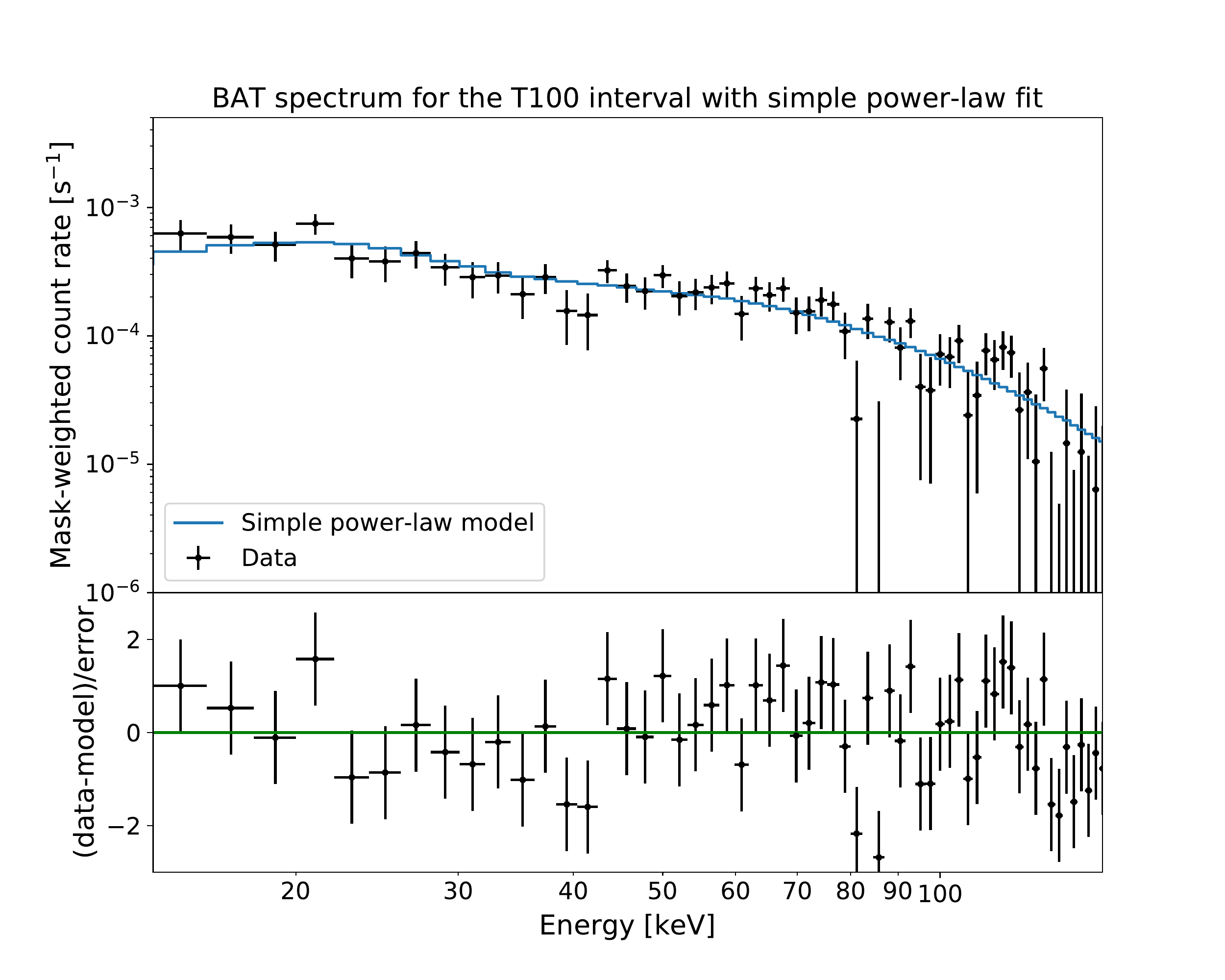}
\caption{{\it Swift}/BAT time averaged spectrum in the $T_{100}$ duration with a simple power-law fit.}
\label{fig:BAT_spectrum}
\end{figure}

The BAT spectral analysis was performed by XSPEC\footnote{ http://heasarc.gsfc.nasa.gov/xanadu/xspec/}. For the spectrum time period that covers the spacecraft slew time (up to $143$ s after the trigger), an average response file is generated by summing up several short-interval (five-second) response files (see detailed description in Lien et al. 2016). 
The time averaged spectrum from T$_0-42.2$ s to T$_0+197.8$ s (the $T_{100}$ duration) can be fitted by a simple power-law model with reduced $\chi^2$ = 1.08 for 57 degrees of freedom (dof). The power law index of the time-averaged spectrum is 1.37 $\pm$ 0.14. The cutoff power-law model gives a reduced $\chi^2$ = 1.09 for 56 degrees of freedom (d.o.f.), and hence does not show significant improvement over the simple power-law model. Fig.~\ref{fig:BAT_spectrum} shows the spectrum in the $T_{100}$ duration and the simple power-law fit.


\subsubsection{Konus-WIND data reduction and analysis}

The Konus-WIND instrument (KW, Aptekar et al. 1995) is a $\gamma$-ray spectrometer consisting of two identical NaI(Tl) detectors, S1 and S2, which observe the southern and northern ecliptic hemispheres, respectively. Each detector has an effective area of 80-160 cm$^{2}$, depending on the photon energy and incident angle. Since KW did not trigger on the burst, the data are available only from the instrument’s "waiting mode". In this mode, count rates with a coarse time resolution of 2.944 s are recorded in three energy bands: G1(22-95 keV), G2(95-390 keV), and G3(390- 1580 keV).

\begin{figure}[!h]
\includegraphics[width=\columnwidth]{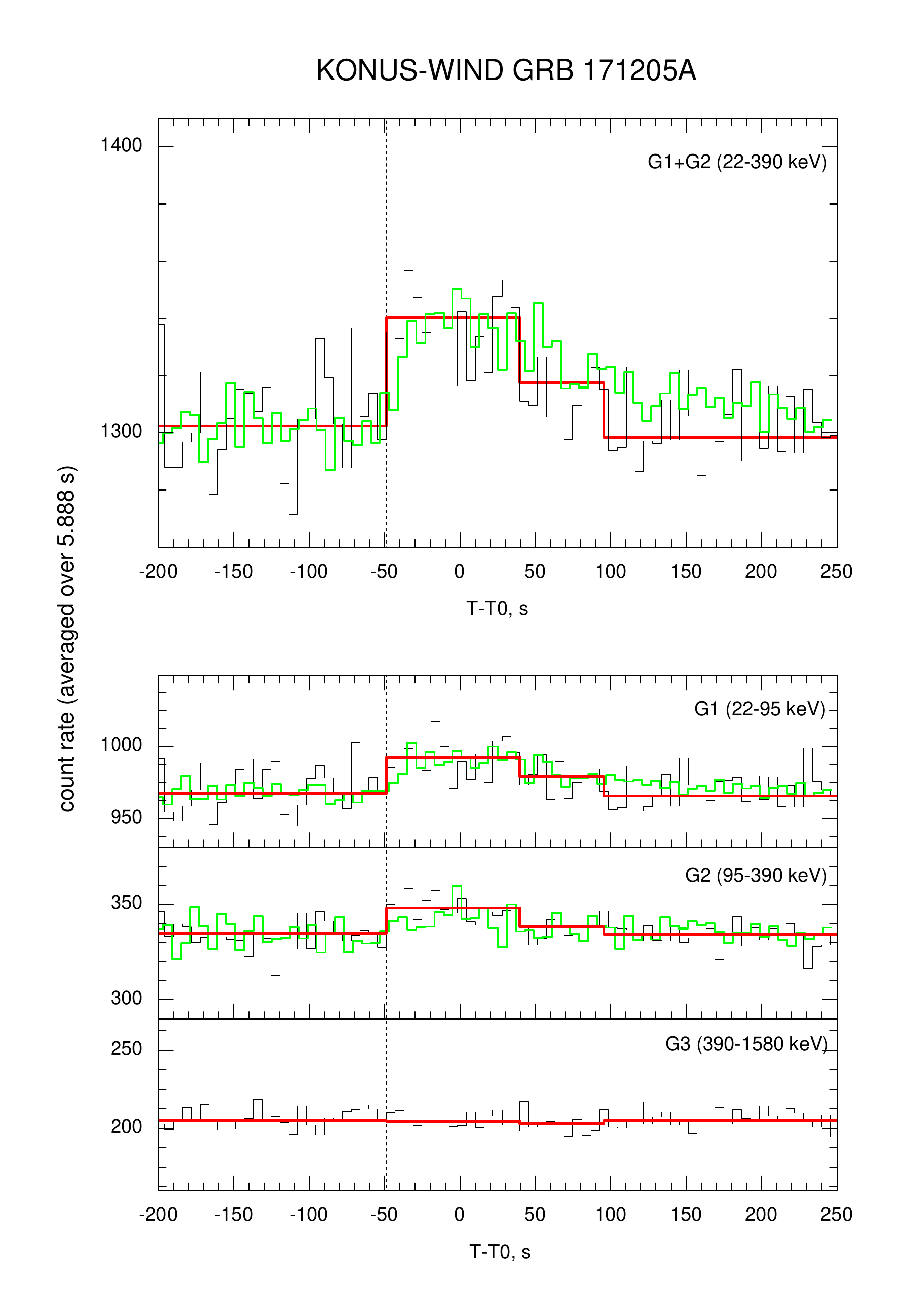}
\caption{Konus-{\it WIND} observation of GRB 171205A in G1+G2, G1, G2, and G3 bands (top to bottom, black lines). Red lines show the results of the Bayesian block analysis in the G1 + G2 band. The Swift-BAT light curves in corresponding energy bands are shown for reference (green lines), shifted and scaled to match the KW light curves. The vertical dashed lines denote the data interval chosen for the joint KW+BAT spectral analysis.}
\label{fig:KW_lc}
\end{figure}

A Bayesian block analysis of the KW waiting mode data reveals a weak count rate increase in S1 in the interval from T$_0-50$ s to T$_0+95$ s. The burst detection significance in the combined G1+G2 light curve is $\sim$ 10$\sigma$ (95 s scale), while no statistically significant emission has been detected in the G3 band (Fig.\ref{fig:KW_lc}).
The time-averaged three-channel spectrum of the most intense part of the burst, measured from T$_0-47.797$ s to T$_0+41.575$ s, is well fit by a simple power-law (PL) model with the PL index 2.00 $\pm$ 0.18, $\chi^2$ = 0.88 for 1 dof.

\subsubsection{Konus-WIND and Swift-BAT joint spectral analysis}

To derive the broad-band spectral parameters of this burst, we performed joint spectral analysis of the {\it Swift}/BAT data and the Konus-Wind three-channel spectral data.
The energy ranges which we used in the joint spectral analysis are 22-1580 keV and 15-150 keV for the KW and the BAT, respectively. The spectral data of two instruments were fitted in XSPEC with the spectral model multiplied by the constant factor to take into account the systematic effective area uncertainties in the response matrices of each instrument. The resulting BAT constant factor $C_\mathrm{BAT}$ (the KW constant factor is fixed to one) is consistent, within uncertainties, with the range (0.8--1.0) derived from the KW-BAT cross-calibration (Sakamoto et al. 2011).

The time interval of the spectral data for each instrument is chosen from T$_0-47.797$ to T$_0+96.459$ s (to ensure the same spectrum accumulation interval, the KW time has been corrected for the burst propagation from Konus-WIND to {\it Swift}). This interval comprises whole burst as observed by Konus-Wind and contains 80\% of the total count fluence in the 15-150 keV as observed by BAT.

\begin{figure}[!h]
\includegraphics[width=\columnwidth]{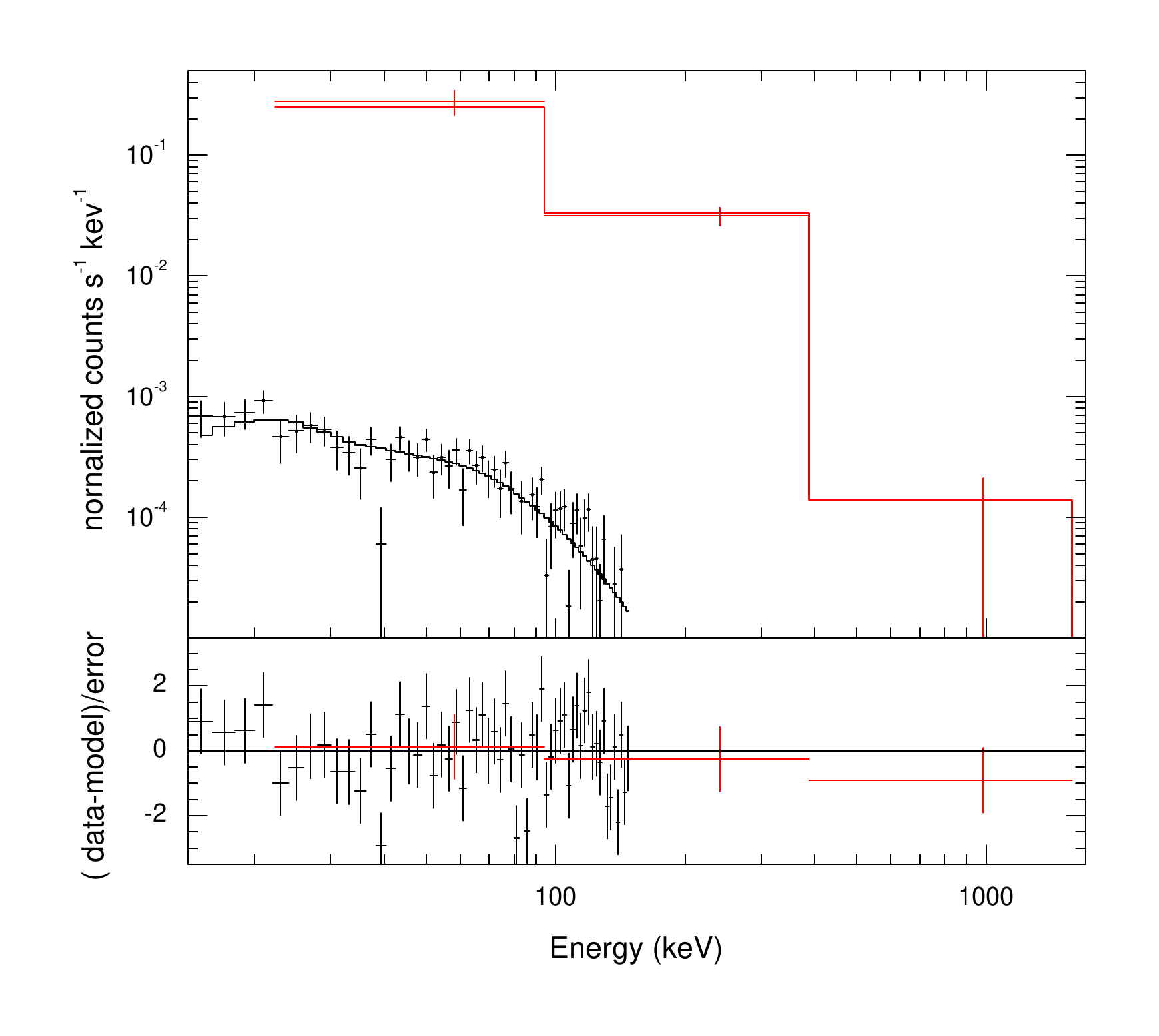}
\caption{Konus-{\it WIND} (red points and line) and {\it Swift}-BAT (black points and line) joint spectral fit to the time-averaged spectrum with the CPL model.}
\label{fig:KW_lc_spec}
\end{figure}

A fit to the spectrum with a simple power-law function results in a photon index $\alpha = 1.50_{-0.14}^{+0.14}$ with a reduced $\chi^2 = 1.48$ for $59 dof$.
The reduced $\chi^2$ gives the best agreement adopting as a model a power-law with exponential cutoff (CPL). No systematic residual from the best fit model is seen in the spectral data of each instrument (Fig. \ref{fig:KW_lc_spec}). The best-fit spectral parameters are: $\alpha = -0.85_{-0.41}^{+0.54}$) and E$_{\rm peak} = 122_{-32}^{+111}$ keV (reduced $\chi^2 = 1.24$ for $58$ dof).

A fit to this spectrum with the Band GRB function yields the same $\alpha$ and E$_{\rm peak}$ and only an upper limit on $\beta < -2.2$ (reduced $\chi^2= 1.26$ for $57$ dof). The best fit spectral parameters for the Band model fixing $\beta$ to $-2.5$ are: $\alpha = -0.88_{-0.42}^{+0.57}$ and E$_{\rm peak} = 125_{-37}^{+141}$ keV (reduced $\chi^2 = 1.27$ for $58$ dof). 

To conclude, despite large uncertainties, we are able to constrain both the lower ($\sim90$ keV) and the upper ($\sim230$ keV) boundary on E$_{\rm peak}$ thanks to the use of the {\it Swift}/BAT and KW data, respectively, the synergy of the two instruments being the key of the success here. Concerning the spectral model used, assuming a Band model instead of a CPL has little impact on the result, either if we freeze $\beta$ to $-2.5$ or if we adopt the hardest spectrum allowed by our data ($\beta < -2.2$, see Table \ref{tab:KW_BAT_spec_fit}).

\begin{table*} \centering
\caption{Results of the spectral fit of KW and BAT data from time interval T$_0-47.797$ to T$_0+96.459$ s.}
\label{tab:KW_BAT_spec_fit}
\begin{tabular}{|cccccc|}
\hline\hline
Instrument &  Model & $\alpha$  & $\beta$ & $E_{\rm peak}$ &  $\chi^2$ (d.o.f) \\
                        &               & (PL index)  & (keV) & & \\
\hline
BAT    & PL   & $ 1.36_{-0.15}^{+0.15}$ & ---            & ---                               & 70.9 (57)\\
\hline
BAT    & CPL  & $-1.08_{-0.37}^{+0.65}$ & ---            & $>94$                          & 70.1 (56)\\
\hline
KW     & PL   & $ 2.08_{-0.29}^{+0.31}$ & ---            & ---                          & 3.9 (1)\\
\hline
        KW     & CPL  & $-0.96$           & ---            & $128$                & 0 (0)\\
\hline
BAT+KW & PL   & $ 1.50_{-0.14}^{+0.14}$ & ---            & ---                 &  87.6 (59)\\
\hline
BAT+KW & CPL  & $-0.85_{-0.41}^{+0.54}$ & ---            & $122_{-32}^{+111} $ &  71.8 (58)\\
\hline
BAT+KW & Band & $-0.85_{-0.41}^{+0.54}$ & $<-2.2$        & $123_{-32}^{+111} $ &  71.8 (57)\\
\hline
BAT+KW & Band & $-0.88_{-0.42}^{+0.57}$ & $-2.5$ (fixed) & $125_{-37}^{+141} $ &  73.3 (58)\\
\hline
BAT+KW & Band & $-0.91_{-0.41}^{+0.32}$ & $-2.2$ (fixed) & $128_{-42}^{+170} $ &  74.6 (58)\\
\hline\hline
\end{tabular}                     
\end{table*}

\subsection{XRT data reduction and analysis}

Observations of  GRB\,171205A by \emph{Swift}/XRT were begun about 150 s after the 
BAT trigger time ($T_{\rm 0}$\,=\,2017-12-05 07:20:43 UT); in this first orbit \emph{Swift}
collected 9.5 s of data as the satellite was slewing to the target, and 250 s 
of pointing data in Windowed Timing (WT) mode (D'Elia et al. 2017).
\emph{Swift} then regularly monitored the afterglow until 2018 February 21 in PC mode for a total exposure time of $\sim$\,305 ks. 
Table~\ref{obs-log} shows the log of all the \emph{Swift} XRT observations. 

The data were processed using the XRTDAS software (v. 3.4.0, Capalbi et
al. 2005) developed at the ASI Space Science Data Center and included in the
HEAsoft package (v. 6.22.1) distributed by HEASARC. For each observation of
the sample, calibrated and cleaned WT and PC mode event files were produced
with the {\it xrtpipeline} task. The default screening criteria were applied to the data.

\begin{figure}[!h]
\includegraphics[width=\columnwidth,angle=0]{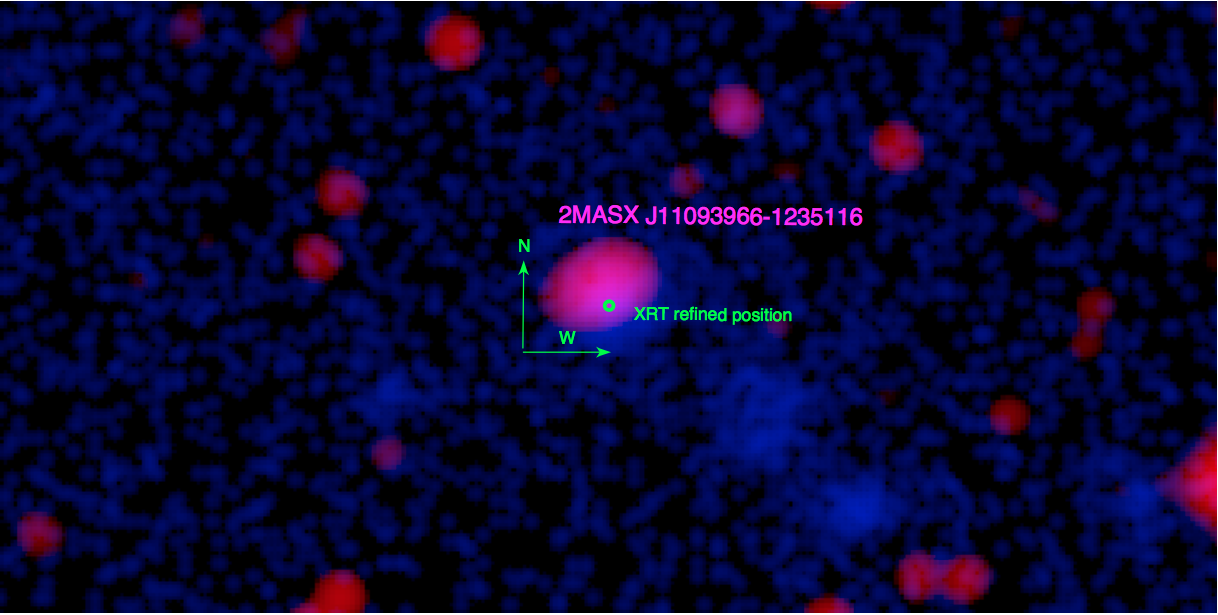}
\caption{{\it Swift}-XRT (0.3--10 keV) data in blue (logarithmic scale) superposed on the optical image from the Sloan Digital Sky Survey in red color.
Data are smoothed with a Gaussian kernel of 3 arcsec width and 1.5 arcsec sigma. 
North and west directions shown by the two arrows of 30 arcsec lenght. 
The refined XRT position has a radius of 1.5 arcsec and it clearly indicates an offset 
$\sim$\,8 arcsec from the center of the host galaxy 2MASX J11093966-1235116.}
\label{fig:xrtimage}
\end{figure}

For PC (WT) data, we extracted  high-level scientific products using a
circular (strip) region  of  20 pixel  radius  (one  pixel\,=\,2.36\,\arcsec)
centered   on source
in the 0.3--10 keV energy range.
We choose as the background a region of similar area
next to the afterglow position where no other contaminating
source is present. In Fig.~\ref{fig:xrtimage}  we show the field of
GRB\,171205A, obtained superposing the XRT refined position to the optical image from the Sloan Digital Sky Survey. 

\begin{figure*}[!h]
\includegraphics[width=19cm]{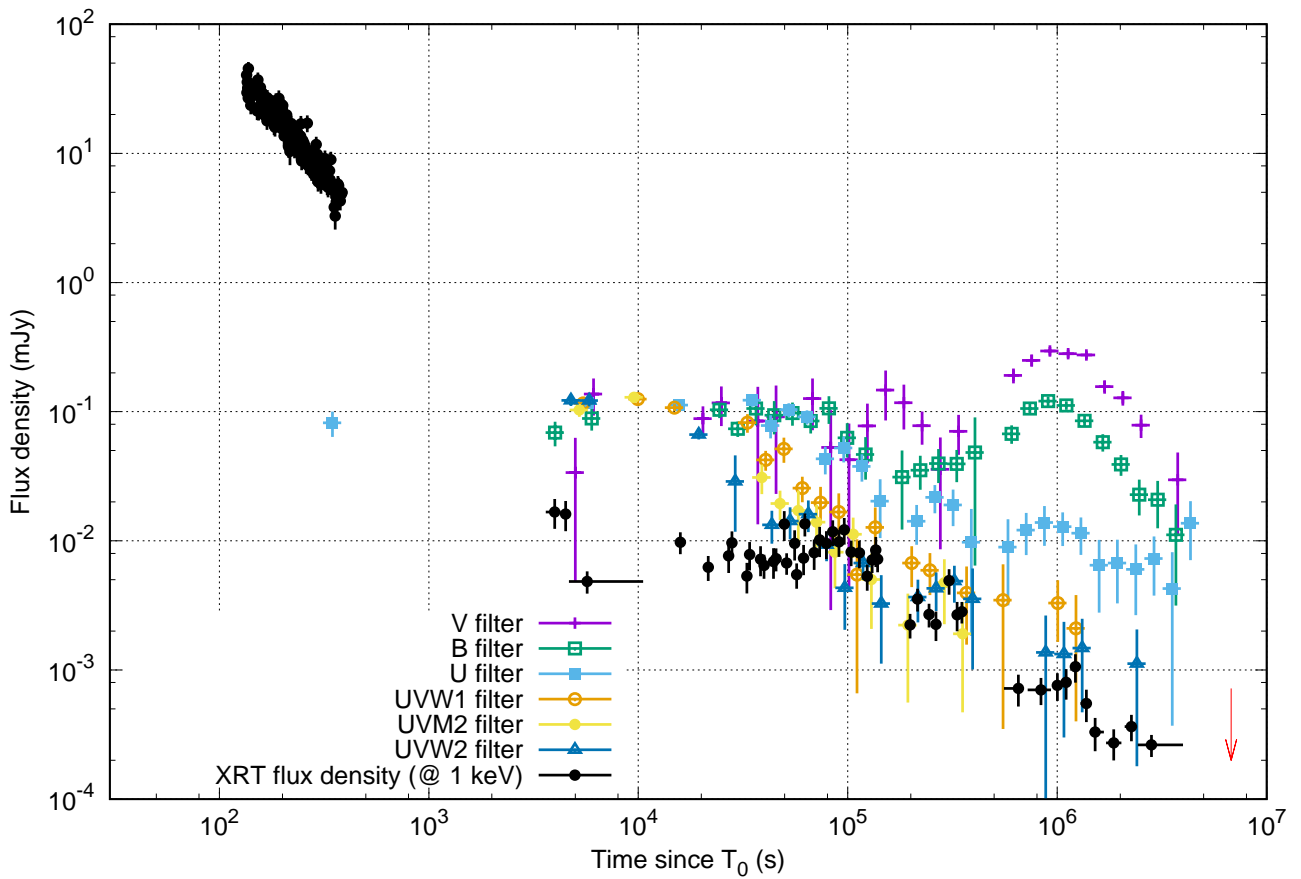}
\caption{{\it Swift}/XRT and /UVOT light curve of the GRB\,171205A afterglow. UVOT data are corrected for the host galaxy and extinction contribution. Information on the UVOT filter passbands can be found at: http://www.swift.ac.uk/analysis/uvot/filters.php.}
\label{fig:lcurve}
\end{figure*}

The X-ray light curve of the afterglow can be well described by a 
steep (index 2.2) power-law from $T_{\rm 0}$+300 s to $T_{\rm 0}$+7.2 ks,
followed by a plateau phase (index $\sim$ 0) from  $T_{\rm 0}$+7.2 ks to 
$T_{\rm 0}$+92 ks, followed by a power-law decay of index 1.08, in other words, a typical steep-flat-normal behavior (Kennea et al. 2017). The data before $T_{\rm 0}$+300 s are described by a shallower power-law (index 1.8) with respect to the steep decay. The XRT light curve is reported in Fig.~\ref{fig:lcurve}.


%
%

\subsubsection{Spectral analysis}

\begin{figure}[!h]
\includegraphics[width=\columnwidth, angle=-0]{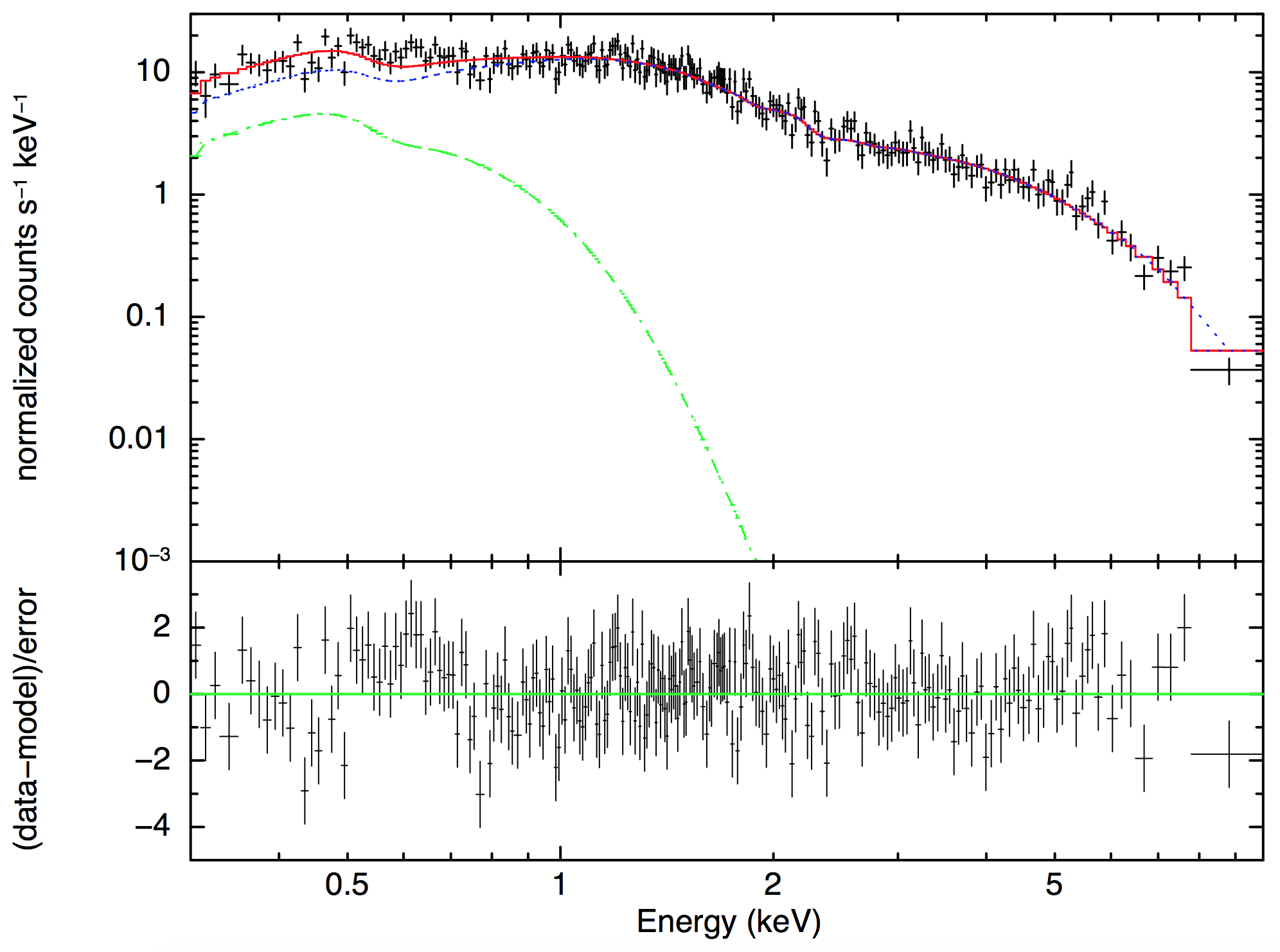}
\caption{Top panel: {\it Swift}/XRT (0.3--10 keV) unfolded spectrum and best-fit model (blackbody component in green and power-law in blue). Bottom panel: Residuals of the fit to the data.}
\label{fig:spectrum}
\end{figure}

During the first XRT observation performed in WT mode, from 150 s to 400 s after the BAT trigger, 
the residuals in the softest XRT band suggested the presence of an additional thermal component
(Campana et al. 2017). 
%
%
%
%
We divided the XRT spectral data into two datasets each containing an equal number of photons (splitting the observation at $\sim 225$ s after trigger).
We then fit the two datasets to investigate spectral changes, but we find compatible results between the two datasets. 
To better investigate the presence of a soft component we included in the fit also the 45 s of BAT data contemporaneous with the XRT observations. We fit the BAT+XRT data with a power law plus a blackbody component plus a Galactic absorption component (fixed to $5.9\times 10^{20}$ cm$^{-2}$, Willingale et al. 2013, modeled with {\tt tbabs}). The soft blackbody component is significant based on an F-test (F statistic value $= 19.2608$ and probability of null hypotesis $= 1.7 \times 10^{-8}$.
We also included an intrinsic absorption component at the redshift $z$=0.0368 of the host galaxy (Izzo et al. 2017b). The inclusion of this additional absorption component is significant at $0.4\%$ (F-test). The overall fit is good, with reduced $\chi^2$=0.99 for 246 dof. The blackbody temperature is $89^{+13}_{-9}$eV and its radius at 163 Mpc is $1.5^{+1.2}_{-0.7}\times 10^{12}$ cm. The intrinsic absorption is $9^{+6}_{-5}\times 10^{20}$ cm$^{-2}$. 
The power law photon index is $\Gamma=1.64\pm0.07$. The mean 0.3--10 keV unabsorbed flux during the WT observation is  $1.5\times 10^{-9}$ erg cm$^{-2}$ s$^{-1}$ (corresponding to a luminosity of $4.8\,10^{45}$ erg s$^{-1})$, with the blackbody component comprising $\sim 20\%$.  We show in Fig.~\ref{fig:spectrum} the XRT unfolded spectrum of the best fit model, including the blackbody component, together with the residuals. 

\begin{figure}[!h]
\includegraphics[width=\columnwidth]{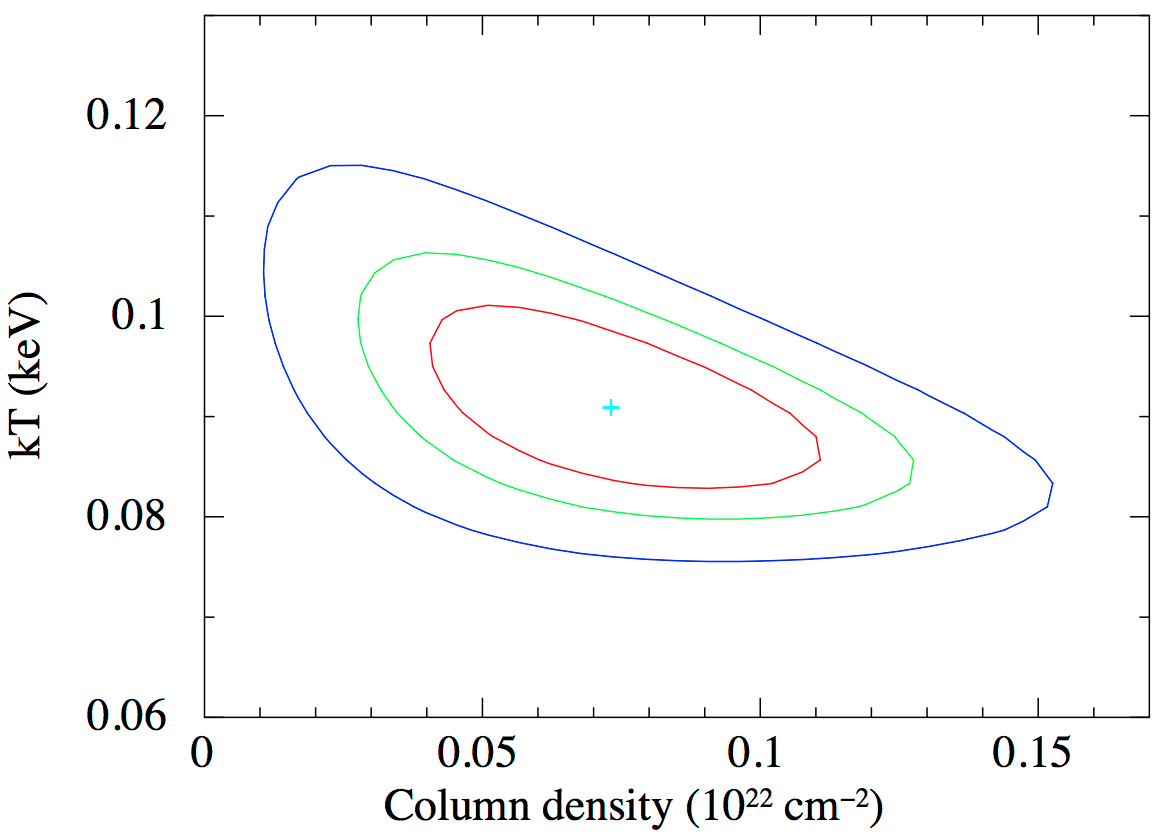}
\caption{Contour plot showing intrinsic column density vs. temperature for the combined WT and late time PC spectrum. Red, green and blue curves enclose $1\sigma$, $2\sigma$ and $3\sigma$ confidence contours, respectively.}
\label{fig:contour}
\end{figure}

We time-selected the PC-mode data and extracted energy spectra in correspondence with the time intervals having the same decay index in the light curve. We adopted a power law spectrum, imposing the same absorption column on the BAT+WT-mode and PC-mode spectra and we found no significant change in slope of the power-law ($Gamma$=$1.83\pm0.06$). We therefore averaged all the data in PC mode and made a common fit with the WT data. The intrinsic host-galaxy absorption is further constrained to be $7.4^{+4.1}_{-3.6}\times 10^{20}$ cm$^{-2}$ at the host redshift of $z=$0.0368 (see Fig. \ref{fig:contour}). The parameters of the (WT) blackbody changed to $91^{+11}_{-9}$eV for the temperature, and $1.3^{+0.8}_{-0.5
}\times 10^{12}$ cm for the radius at $163$ Mpc ($\Gamma=1.63\pm0.06$), that is, they are consistent within the errors. The  blackbody component is still comprising $\sim 17\%$ of the 0.3--10 keV flux. 
The overall fit is good with reduced $\chi^2=0.97$ for $284$ dof. To conclude, we note the intrinsic X-ray column density we obtain is at the low end of the GRB distribution, even among low redshift {\it Swift}-XRT GRBs where the mean is  $N_{H,int} \sim 2.4 \times 10^{21}$ cm$^{-2}$ at $z < 0.2$ (Arcodia, Campana \& Salvaterra 2016).

\subsection{UVOT data reduction}

The Swift/UVOT began observing the field of GRB 171205A 154 s after the trigger (D'Elia et al. 2017). The afterglow was detected in all UVOT filters.

The source counts were extracted using a region of $3''$ radius. We used this aperture size because whilst the standard aperture for extracting count rates is a $5''$ radius aperture, it is more accurate to use a smaller source aperture if the count rate is low (Poole et al. 2008). Furthermore, a smaller aperture limited the contamination of the host galaxy. The count rates were corrected to a $5''$ aperture using the curve of growth from the UVOT calibration files so that they are consistent with the UVOT calibration. Background counts were extracted using a circular region of radius $30''$ in a source-free region where the is no substantial contribution from the host galaxy. The count rates were measured using the Swift ftool {\it uvotsource} and were converted to magnitudes using the UVOT photometric zero points (Breeveld et al. 2011). The UVOT data are provided in Table 3 in the appendix. The analysis pipeline used software HEADAS 6.21 and UVOT calibration 20170130.

Since we observed GRB 171205A superposed on its host galaxy, the count rates we measured are contaminated with light from the host. To measure the contribution of the host galaxy, we measured the count rates for each observation between 6729--6788 ks using the same $3''$ source aperture we used for GRB 171205A and performed a moving mean calculation to get an average count rate for the host galaxy. The host galaxy count rates were aperture-corrected to $5''$ for the same reasons as the source count rates were. Host galaxy correction was performed on all count rates for each filter.

For each light curve the magnitudes were binned with $\Delta t/t=0.2$, when the signal-to-noise ratio, $(S/N) < 2$. Additionally, the light curves were adjusted to correct for Galactic extinction, $E(B-V)=0.0434$ (Schlafly \& Finkbeiner 2011). The resulting background subtracted, galactic extinction corrected light curves are displayed in Fig.~\ref{fig:lcurve}. No correction for the host galaxy absorption was performed.

\section{Discussion}

\subsection{$\gamma$-ray energetics and spectral shape}

Using the joint BAT and Konus-Wind spectral fit, the burst energy fluence in the 15-1500 keV band calculated by the best-fit (CPL) model for the 144.26 s interval is $S = 5.98^{+1.8}_{-1.41} \times 10^{-6} \ \rm erg \ cm^{-2}$. Assuming $z=0.0368$ (Izzo et al. 2017b), we estimated the isotropic energy release (1 keV to 10 MeV in the GRB rest frame) of GRB 171205A to be $E_{\rm iso} = 2.18^{+0.63}_{-0.50} \times 10^{49}$ erg. 

In this approach, the $E_{\rm iso}$ errors are propagated directly from the flux errors, that is, not taking into account the k-correction uncertainty. We checked that, calculating the model fluence in the bolometric rest-frame band to take this effect into account, increases just slightly the $E_{\rm iso}$ error (less than $10\%$). Concerning the spectral model uncertainties, if we use a Band function instead of a CPL, we obtain $E_{\rm iso} = 3.6^{+0.9}_{-0.9} \times 10^{49}$, a slightly higher value. 

\begin{figure}[!h]
\includegraphics[width=\columnwidth]{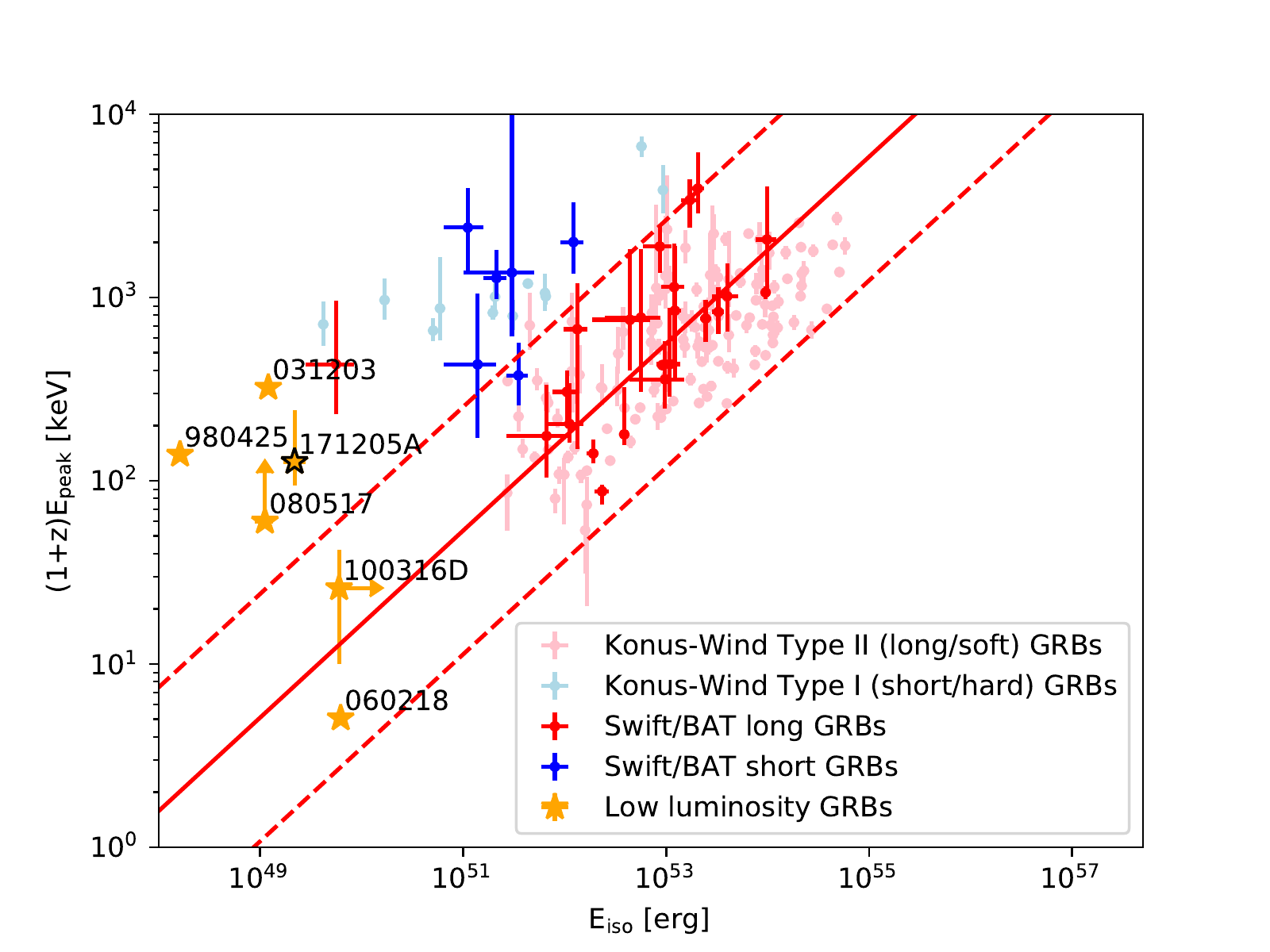}
\caption{$E_{\rm peak}$ in rest-frame versus $E_{\rm iso}$. The {\it Swift}/BAT GRB sample (dark blue and red) is adapted from Krimm et al. (2009). The Konus-Wind GRB sample (light blue and red) is adapted from Tsvetkova et al. (2017). The yellow stars show GRB 171205A and several sources defined as low-luminosity GRBs in previous studies (Campana et al. 2006; Ghisellini et al. 2006; Starling et al. 2011; Stanway et al. 2015). The red lines are the best fit (solid line) and the 2.5$\sigma$ variation (dashed line) reported in Krimm et al. (2009).}
\label{fig:Epeak_Eiso}
\end{figure}

The $E_{\rm iso}$ derived for this burst is similar to those of other low-luminosity GRBs. For example, the $E_{\rm iso}$ (in the observed 15-150 keV band) for GRB 060218 is $2.57 \times 10^{49}$ erg (assuming $T_{90}$ = 2100 s, see Campana et al. 2006) and GRB 100316D has $E_{\rm iso} \geq 3.70 \times 10^{49}$ erg (assuming $T_{90} \geq 1300$ s, see Starling et al. 2011). However, besides the low $E_{\rm iso}$, other properties of GRB 171205A seem to be different from GRB 060218 and GRB 100316D. The BAT spectrum of GRB 171205A is harder than those of GRB 060218 and GRB 100316D, which have power-law indices of $2.18$ and $2.36$, respectively (Lien et al. 2016). In addition, both GRB060218 and GRB 100316D show long-lasting burst emission of more than $\sim 1000$ s, while GRB171205A has no obvious emission after $\sim$ T$_0+200$ s, though we note that GRB171205A went out of the BAT field of view at T$_0+479$ s. 
As shown in Fig.~\ref{fig:Epeak_Eiso}, GRB 171205A lies outside the Amati relation from the BAT long GRB (Krimm et al. 2009) and the Konus-Wind (Tsvetkova et al. 2017) samples, even adopting the more conservative values for E$_{\rm peak}$ (see Section 3.1.3) and $E_{\rm iso}$.

%
%



Other low-luminosity GRBs such as 171205A are not consistent with the Amati relation. As discussed by Amati et al. (2007), this lack of consistency strongly indicates that these LL-GRBs are not just normal, "cosmological" events seen off-axis. GRBs are supposed to be jetted sources seen on-axis (Sari et al. 1999) or very close to it. Theoretically, if a GRB event were seen off-axis, the prompt emission detected would be weaker, since the radiation is strongly beamed toward the direction of motion by relativistic effects. It is found that $E_{\rm p} \propto \delta$ and $E_{\rm iso} \propto \delta^{1-\alpha}$ (Amati et al. 2007, Yamazaki et al. 2003), where $\alpha$ is the low-energy photon index and $\delta$ is the relativistic Dopper factor which, in turn, depends on the half-opening angle $\theta_{jet}$ of the GRB jets and the angle between the GRB jet axis and the observer $\theta_{obs}$: $\delta = (\Gamma(1 - v/c {~\rm cos}(\theta_{obs} - \theta_{jet})))^{-1}$.  This parameter $\delta$ decreases as $\theta_{obs} - \theta _{jet}$ increases.


Since E$_{\rm p}$ and E$_{\rm iso}$ depend on $\delta$ in different ways, GRBs seen off-axis cannot follow the Amati relation as GRBs seen on-axis. We find that, in principle, this might be the case for 171205A. Let us assume $\alpha = -1.2$, which is still consistent with the analysis of the prompt emission at 90\% confidence level.  Let us also assume that  this GRB is seen off-axis and that  $\delta$ decreases by a factor of $\sim 500$  from when this event is observed on-axis. Thus, the estimate of E$_{\rm p}$ would decrease by $\sim 500$, while the estimate of E$_{\rm iso}$ would decrease by $\sim500^{1+1.2} = 8.7\times10^5$.  The on-axis values of these parameters would thus be E$_{\rm p,onaxis} \sim 6\times10^4$~keV and E$_{\rm iso,onaxis} \sim 2\times10^{56}$~erg. In principle, these estimates are now within $\sim 2.5\sigma$ variation from the best fit of the Amati relation. However, these values are also highly problematic because they are large compared to those of known GRBs. A GRB with such parameters would be truly exceptional and unlikely to be found within the relatively small volume enclosed by the redshift of GRB 171205A. A higher value of $\alpha$ would only exacerbate the problem, while lower values of $\delta$ would not allow the parameters of this GRB to be consistent with the Amati relation. 
Thus, GRB171205A  seems to be an event that cannot be explained as a typical cosmological event seen off-axis; instead, its emission mechanism appears to be different from those of farther away, very energetic GRBs.

We conclude the section with a word of caution on the outliers of the Amati relation. The relation below $10^{50}$ erg is not well studied
yet, and in some cases the locations of the outliers could be due to observational biases (Martone el al. 2017). In addition, the upper boundary of the Amati relation,
as distinct from the lower one, is strongly affected by instrumental
selection effects (Heussaff et al. (2013); 
Tsvetkova et al. (2017)) and could not be unequivocally treated as an intrinsic
GRB property. Thus, the problem of the upper-side outliers in the Amati
relation, especially at low E$_{\rm iso}$, is rather complicated.

\subsection{The UV-optical light curve}


\begin{figure}[!h]
\includegraphics[width=\columnwidth]{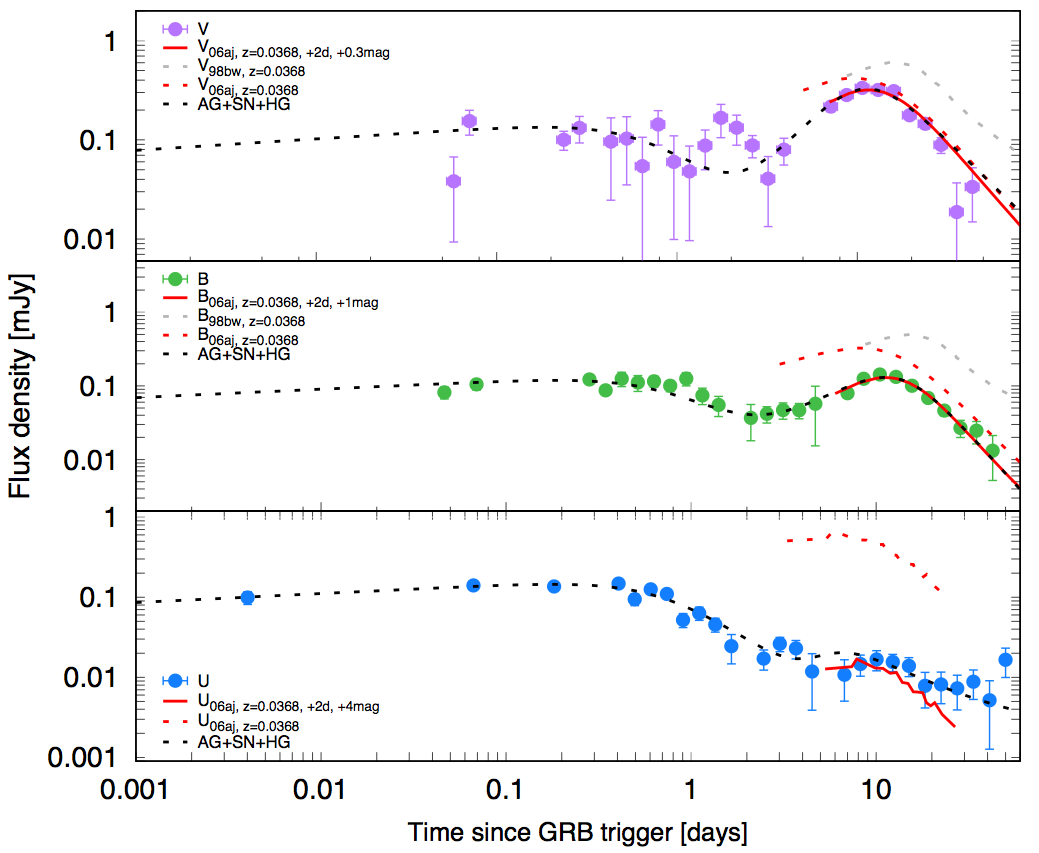}
\caption{Overall fit of the $ubv$ light curves (dashed black line). The best match for the SN contribution is obtained with the SN\,2006aj template (dashed red lines) shifted by +2 days but lower in magnitude (solid red lines).}
\label{fig:sn2017iuk}
\end{figure}

At early times, the UV-optical light curves (Fig.~\ref{fig:lcurve}) are flat ($\alpha_{\rm 1,U}$ = -0.11 $\pm$ 0.08), consistent with the plateau phase observed in the X-rays, followed by a steeper decay phase ($\alpha_{\rm 2,U}$ = 1.79 $\pm$ 0.39). This behaviour has been previously observed for several {\it Swift} GRBs (i.e., Oates et al. 2009; Melandri et al. 2014). At late times ($\Delta t > 3$ d) the signature of an emerging supernova component is clearly visible in the UBV optical filters, and this was also reported by independent spectroscopic observations (de Ugarte Postigo et al. 2017a).

In Fig.~\ref{fig:sn2017iuk} we fit the early $u$-band light curve (which is the best sampled filter since the start of UVOT observations) with a broken power-law in order to estimate the afterglow contribution. The best fit is then rigidly shifted to the $b$ and $v$-band data. The agreement with the data is good and therefore we can consider the UVOT afterglow to be achromatic. Then we compare the observed $u$, $b$, and $v$ light curves for SN\,2017iuk with the corresponding curves for SN\,1998bw and SN\,2006aj. The best match of the SN bump is with the template light curves of SN\,2006aj, but lower in magnitude and with the peak time shifted by $\sim$2 d. A simple estimate of the peak time for the $b$ and $v$ filters provides the following values: t$_{\rm peak}^{\rm B}$ = 12.7 $\pm$ 1.2 and t$_{\rm peak}^{\rm V}$ = 13.6 $\pm$ 1.5 days. These values are consistent with typical peak times and magnitudes of other well-studied SNe connected with low-luminous GRBs (i.e., Cano et al. 2017a). Finally, we summed the afterglow, supernova, and host galaxy contribution obtaining the overall fit of the light curves in the UBV filters (Fig.~\ref{fig:sn2017iuk}).

The description of the properties of SN\,2017iuk is beyond the purpose of this paper and it would require a more detailed modeling and subtraction of the afterglow contribution in each UVOT filter. However, we must note that despite the fact that the behaviour of the UV-optical light curve at $t<5$ d seems to be achromatic, it displays a possible premaximum bump in the $v$-band that is not visible in other filters. As discussed by Cano et al. (2017b) such early bumps have been observed for other GRBs-SNe and they typically seem to be achromatic; for SN\,2017iuk (as for SN\,2016jca, discussed by Cano et al. 2017b) the bump is detected only in one filter and its physical origin could not be easily explained.

\subsection{X-ray spectral behaviour}


Although the X-ray spectrum is well described by an absorbed power law, the inclusion of a thermal component to fit the data significantly improves the reduced $\chi^2$ (see Section 3.2). The presence of such a component is not new in GRB spectra, especially in the early time data of GRBs associated with SNe, with notable examples being GRB\,060218 (e.g., Campana et al. 2006), GRB\,090618 (e.g., Page et al. 2011) and GRB\,100316D (e.g., Starling et al. 2011). 

The origin of this thermal feature in GRBs is still under debate, with one of the first proposed explanations being the shock breakout of the emerging supernova (Colgate 1974; Waxman, M\'esz\'aros \& Campana 2007; Nakar \& Sari 2010; Nakar \& Sari 2012). Alternative explanations include late jet photospheric emission (e.g., Friis \& Watson 2013), or that the emission comes from a relativistically expanding hot plasma cocoon surrounding the GRB jet (e.g., Suzuki \& Shigeyama 2013; De Colle et al. 2018). In fact, Pe'er, M\'esz\'aros \& Rees (2006) demonstrate that a few hundreds of seconds after the prompt emission, the bulk of the cocoon radiation may fall in the X-ray band.

The first attempt to study the X-ray thermal emission in GRB-SNe from a statistical point of view was performed by Starling et al. (2012). These authors select $11$ {\it Swift} GRBs with associated SNe and systematically searched for a thermal (blackbody-like) emission in the X-ray spectra. They found that four objects do (and  three do not) show this component, and identified a further four possible cases of thermal emission. The conditions on redshift and absorbing column density required for detection of a thermal X-ray component similar to those seen previously in GRB afterglows are assessed in Sparre \& Starling (2012) and are all met in the case of GRB\,171205A. We therefore compared the physical parameters of the Starling et al. sample with those evaluated for our GRB. The ratio between the blackbody and the total 0.3--10\,keV X-ray flux as well as the radius of the blackbody emitting region for GRB\,171205A fall within the corresponding parameter distributions reported in Starling et al. (2012). On the other hand, our GRB has a blackbody temperature that is lower than those of all GRBs reported in Starling et al. (2012) over the first few hundred seconds since burst (thanks also to the closeness of the GRB), but is consistent with their later ($\sim$760\,s) detections of the blackbody in GRB\,101219B .

\section{Conclusions}

The burst GRB 171205A revealed itself as a close-by burst ($z=0.0368$), thus belonging to a class of GRBs comprising few but high flux events. Together with the discovery of the associated supernova (SN 2017iuk), this made GRB 171205A one of the most followed-up bursts of recent years.

In this paper we presented the {\it Swift} and Konus-Wind data of this burst. The dataset extends up to $\sim 10^7$ s in the time domain and from the {\it Swift}-UVOT $v$ band ($\sim 2.25\times 10^{-3}$ keV) to the highest Konus-Wind band ($\sim 1.6 \times 10^3$ keV).

Our main findings are summarized below.
\begin{itemize}
  \item We derive the rest frame peak energy E$_{\rm p}=125^{+141}_{-37}$ keV and isotropic energy $E_{\rm iso} = 2.18^{+0.63}_{-0.50} \times 10^{49}$ erg by fitting the {\it Swift}-BAT and Konus Wind spectral data. As other low luminosity GRBs, GRB 171205A lies outside the Amati relation. This behaviour strongly indicates that this GRB is not a ``normal" event seen off-axis, but its emission mechanism should be different from cosmological bursts.
  \item We studied the optical-UV light curve collected by {\it Swift}-UVOT. The $v$, $b$, and $u$ data clearly show a supernova feature  emerging from the fading contribution of the afterglow radiation. The peak of the SN emission is located at $\sim 13$ days after the GRB detection. Comparing SN 2017iuk light curves with templates in the literature, we found that the best match is with SN 2006aj, but lower in magnitude and with the peak time shifted by approximately two days. We finally note a premaximum bump in the $v$-band. Such features have been observed in other GRB-associated SNe, but not in just a sinlge band as in GRB 171205A (i.e., the premaximum bump was achromatic). The chromaticity of the premaximum bump in the GRB 171205A/SN 2017iuk is thus not easily explained.
   \item We produced the X-ray spectrum of GRB 171205A using the {\it Swift}-XRT data. The intrinsic column density lies at the low end of the GRB distribution, even considering just low redshift {\it Swift}-XRT GRBs. We detected a thermal component in the X-ray spectrum. The ratio between the blackbody and the total 0.3--10\,keV X-ray flux as well as the radius of the blackbody emitting region for GRB\,171205A fall within the the corresponding parameter distributions reported in previous works. On the other hand, our GRB has a blackbody temperature that is lower than that of all GRBs with a reported thermal component at early times. The origin of this component is still a matter of debate.

\end{itemize}

\begin{acknowledgements}
The {\it Swift} team would like to devote this paper to the memory of Neil Gehrels. RLCS acknowledges funding from STFC. KLP and JPO acknowledge support from the UK Space Agency. DDF, DSS, and AET acknowledge support from RSF grant 17-12-01378. DNB and AT acknowledge support from NASA contract NAS5-00136. SRO gratefully acknowledges the support of the Leverhulme Trust Early Career Fellowship.
\end{acknowledgements}

\section*{Appendix}
Tables 2 and 3 report the summary of our XRT and UVOT observations, respectively.

\begin{table*}[bp] 
\centering
\begin{tabular}{||l| c|c| r|r|}
\hline

ObsID (MODE) &  $T_{\rm start}$(TT)&$T_{\rm stop}$ (TT)  & Exposure (s) & Count rate/Error ($10^{-3}$ counts s$^{-1}$)\\
                         \hline\noalign{\smallskip}
794972000 (WT) & 2017-12-05 07:23:30 & 2017-12-05 08:21:12 &   250  & (24.6$\pm$0.3)$\times$10$^{3}$\\
794972000 (PC) & 2017-12-05 08:21:14 & 2017-12-05 08:25:04 &   229  & 29$\pm$13\\
794972001 (PC) & 2017-12-05 08:29:14 & 2017-12-05 15:14:52 &  7964  & 19.0$\pm$1.7\\
794972002 (PC) & 2017-12-05 16:19:40 & 2017-12-06 22:40:54 & 14724  & 16.7$\pm$1.2\\
794972003 (PC) & 2017-12-06 00:31:46 & 2017-12-06 22:55:52 & 19968  & 7.3$\pm$1.3\\
794972004 (PC) & 2017-12-07 14:36:22 & 2017-12-07 21:17:53 &  7302  & 7.2$\pm$1.2 \\
794972005 (PC) & 2017-12-07 11:33:22 & 2017-12-07 13:23:52 &  1753  & 8.1$\pm$2.5\\
794972006 (PC) & 2017-12-08 00:10:04 & 2017-12-08 09:59:52 &  7921  & 6.7$\pm$1.1\\
794972007 (PC) & 2017-12-08 19:15:11 & 2017-12-08 21:14:54 &  2040  & 1.3$\pm$0.3\\
794972008 (PC) & 2017-12-09 00:07:23 & 2017-12-09 13:10:53 &  7354  & 7.8$\pm$1.6\\
794972009 (PC) & 2017-12-09 14:47:03 & 2017-12-09 14:47:03 &   759  & 1.9 (3-$\sigma$ u.l.)\\
794972010 (PC) & 2017-12-11 17:54:11 & 2017-12-11 21:09:54 &   702  & 2.4 (3-$\sigma$ u.l.)\\
794972012 (PC) & 2017-12-12 12:32:17 & 2017-12-12 19:24:52 &  7497  & 1.6$\pm$0.6\\
794972014 (PC) & 2017-12-13 01:45:15 & 2017-12-13 17:49:54 &  4747  & 2.0$\pm$0.8\\
794972015 (PC) & 2017-12-13 19:13:05 & 2017-12-13 19:26:55 &   812  & 2.1 (3-$\sigma$ u.l.)\\
794972016 (PC) & 2017-12-14 01:15:54 & 2017-12-14 14:24:53 &  4915  & 2.0$\pm$0.9\\
794972017 (PC) & 2017-12-15 01:12:53 & 2017-12-15 23:51:52 &  5996  & 2.7$\pm$0.8\\
794972018 (PC) & 2017-12-16 01:07:53 & 2017-12-16 23:54:53 &  9607  & 2.0$\pm$0.5\\
794972019 (PC) & 2017-12-17 01:16:25 & 2017-12-17 18:47:52 &  7165  & 1.4$\pm$0.7\\
794972020 (PC) & 2017-12-18 00:48:53 & 2017-12-18 10:38:53 &  5436  & 3.1$\pm$0.9\\
794972021 (PC) & 2017-12-19 00:57:56 & 2017-12-19 16:57:53 &  8480  & 3.0$\pm$0.7\\
794972022 (PC) & 2017-12-20 05:54:20 & 2017-12-20 20:03:53 &  6545  & 2.2$\pm$0.8\\
794972023 (PC) & 2017-12-21 00:42:03 & 2017-12-21 18:35:54 &  8213  & 2.1$\pm$0.6\\
794972024 (PC) & 2017-12-21 19:47:23 & 2017-12-22 11:58:54 &  8121  & 1.3$\pm$0.5\\
794972025 (PC) & 2017-12-23 02:33:21 & 2017-12-23 13:35:51 & 5614   & 1.0$\pm$0.5\\
794972026 (PC) & 2017-12-24 02:22:53 & 2017-12-24 15:03:52 & 8865   & 2.2 (3-$\sigma$ u.l.)\\
794972027 (PC) & 2017-12-25 02:19:28 & 2017-12-25 22:40:52 & 5432   & 1.2$\pm$0.6\\
794972028 (PC) & 2017-12-26 16:02:25 & 2017-12-26 21:05:53 & 5921   & 3.6 (3-$\sigma$ u.l.)\\
794972029 (PC) & 2017-12-27 14:23:32 & 2017-12-27 21:11:52 & 6231  &  3.3 (3-$\sigma$ u.l.)\\
794972030 (PC) & 2017-12-28 03:27:02 & 2017-12-28 08:01:52 & 2295  &  6.4 (3-$\sigma$ u.l.)\\
794972031 (PC) & 2017-12-29 14:11:03 & 2017-12-29 22:41:54 & 8453  &  2.8 (3-$\sigma$ u.l.)\\
794972032 (PC) & 2017-12-30 03:14:56 & 2017-12-30 19:28:53 & 9632  &  1.3$\pm$0.5\\
794972033 (PC) & 2017-12-31 00:03:15 & 2017-12-31 21:01:52 & 4028  &  4.5 (3-$\sigma$ u.l.)\\
794972034 (PC) & 2018-01-01 13:54:30 & 2018-01-01 23:47:53 & 9427  &  1.4$\pm$0.6\\
794972035 (PC) & 2018-01-02 09:15:27 & 2018-01-02 20:51:52 & 8788  &  3.0 (3-$\sigma$ u.l.)\\
794972036 (PC) & 2018-01-03 02:49:52 & 2018-01-03 14:19:54 & 9550  &  1.0$\pm$0.5\\
794972037 (PC) & 2018-01-04 10:46:59 & 2018-01-04 18:43:52 & 8231  &  1.0$\pm$0.5\\
794972038 (PC) & 2018-01-05 08:47:19 & 2018-01-05 18:57:53 & 9714  &  0.9$\pm$0.4\\
794972039 (PC) & 2018-01-08 08:30:33 & 2018-01-08 15:08:53 & 7557  &  3.1 (3-$\sigma$ u.l.)\\
794972040 (PC) & 2018-01-10 15:09:37 & 2018-01-10 20:05:54 & 4910  &  4.0 (3-$\sigma$ u.l.)\\
794972041 (PC) & 2018-01-12 09:56:31 & 2018-01-12 09:56:31 & 3753  &  5.1 (3-$\sigma$ u.l.)\\
794972042 (PC) & 2018-01-14 08:11:59 & 2018-01-14 14:48:54 &  4852 & 6.0 (3-$\sigma$ u.l.)\\
794972043 (PC) & 2018-01-16 04:54:20 & 2018-01-16 17:57:54 &  5256 & 4.0 (3-$\sigma$ u.l.)\\
794972044 (PC) & 2018-01-19 04:37:37 & 2018-01-20 09:28:52 &  4105 & 4.6 (3-$\sigma$ u.l.)\\
794972045 (PC) & 2018-02-21 04:36:37 & 2018-02-21 20:58:51 &  8490 & 5.0 (3-$\sigma$ u.l.)\\
\hline 
\end{tabular}
\caption{\label{obs-log}  Log of  the \emph{Swift}/XRT observations
used in this work. The
  columns show the  identification number for each observation
  (ObsID, we omitted leading zeros), 
  the $T_{\rm start}$  and $T_{\rm stop}$  times in  TT, 
  the exposure time, 
  the average count rate (or the 3-$\sigma$ upper limit when no statistically significant, above 2 $\sigma$,
  detection is made) in a source box-radius of 10 pixel centered on
  the source coordinates.}
\end{table*}

\begin{table*} \centering
\resizebox{\textwidth}{!}{%
\begin{sideways}

\begin{tabular}{|l|l| c|c|c|c|c|c|c|c|c|c|c|}

\hline 
 Tmid (s)               &Terr (s)        &AB Mag        &Mag E+ &Mag E-   &AB MagU ($3\sigma$)     &Rate (cts/s)         &RateErr (cts/s)        &Flux (mJy)           &Flux err (mJy)      &FluxUL ($3\sigma$)   &Filter      &S/N    \\
\hline                                                                                                                                             
5000.13320      &500.01330     &20.075  &+2.104 &-0.671   &18.694       &1.3247e-01     &1.1340e-01     &0.03377        &0.02891        &0.12048        &V      &1.168  \\
6111.27390      &611.12740     &18.554  &+0.416 &-0.300   &17.827       &5.3736e-01     &1.7101e-01     &0.13698        &0.04359        &0.26775        &V      &3.142  \\
20371.95930     &2037.19590    &19.030  &+0.305 &-0.238   &18.432       &3.4678e-01     &8.4968e-02     &0.08840        &0.02166        &0.15337        &V      &4.081  \\
24899.06130     &2489.90610    &18.725  &+0.449 &-0.316   &17.964       &4.5931e-01     &1.5543e-01     &0.11708        &0.03962        &0.23594        &V      &2.955  \\
37194.89410     &3719.48940    &19.078  &+1.998 &-0.663   &17.710       &3.3182e-01     &2.7914e-01     &0.08458        &0.07115        &0.29805        &V      &1.189  \\
45460.42610     &4546.04260    &18.997  &+1.492 &-0.606   &17.720       &3.5760e-01     &2.6712e-01     &0.09115        &0.06809        &0.29543        &V      &1.339  \\
55562.74310     &5556.27430    &19.695  &+nan   &-0.798   &18.122       &1.8794e-01     &2.0406e-01     &0.04791        &0.05202        &0.20396        &V      &0.921  \\
67910.01930     &6791.00190    &18.643  &+0.612 &-0.389   &17.742       &4.9503e-01     &2.1336e-01     &0.12619        &0.05439        &0.28935        &V      &2.320  \\
83001.13470     &8300.11350    &19.589  &+3.147 &-0.722   &18.130       &2.0721e-01     &1.9579e-01     &0.05282        &0.04991        &0.20254        &V      &1.058  \\
101445.83130    &10144.58310   &19.826  &+2.556 &-0.700   &18.401       &1.6662e-01     &1.5079e-01     &0.04247        &0.03844        &0.15778        &V      &1.105  \\
123989.34940    &12398.93490   &19.171  &+0.726 &-0.431   &18.193       &3.0438e-01     &1.4839e-01     &0.07759        &0.03783        &0.19107        &V      &2.051  \\
151542.53810    &15154.25380   &18.478  &+0.587 &-0.379   &17.596       &5.7671e-01     &2.4074e-01     &0.14701        &0.06137        &0.33111        &V      &2.396  \\
185218.65770    &18521.86580   &18.722  &+0.519 &-0.350   &17.896       &4.6063e-01     &1.7505e-01     &0.11742        &0.04462        &0.25128        &V      &2.632  \\
226378.35940    &22637.83590   &19.169  &+0.366 &-0.273   &18.496       &3.0519e-01     &8.7246e-02     &0.07780        &0.02224        &0.14451        &V      &3.498  \\
276684.66150    &27668.46610   &20.010  &+1.551 &-0.614   &18.720       &1.4055e-01     &1.0688e-01     &0.03583        &0.02724        &0.11756        &V      &1.315  \\
338170.14180    &33817.01420   &19.277  &+0.455 &-0.320   &18.510       &2.7614e-01     &9.4483e-02     &0.07039        &0.02408        &0.14264        &V      &2.923  \\
617427.24110    &61742.72410   &18.195  &+0.152 &-0.133   &17.836       &7.4825e-01     &9.7565e-02     &0.19073        &0.02487        &0.26534        &V      &7.669  \\
754633.29460    &75463.32950   &17.902  &+0.124 &-0.111   &17.598       &9.7991e-01     &1.0543e-01     &0.24978    &0.0268     &0.33041 &V  &9.294  \\
922329.58230    &92232.95820   &17.722  &+0.085 &-0.079   &17.500       &1.1568e+00     &8.7502e-02     &0.29489        &0.02230        &0.36180        &V      &13.221 \\
1127291.71170   &112729.17120  &17.773  &+0.081 &-0.075   &17.561       &1.1041e+00     &7.8959e-02     &0.28143        &0.02013        &0.34181        &V      &13.983 \\
1377800.98100   &137780.09810  &17.799  &+0.079 &-0.074   &17.591       &1.0772e+00     &7.5863e-02     &0.27459        &0.01934        &0.33261        &V      &14.200 \\
1683978.97680   &168397.89770  &18.411  &+0.138 &-0.123   &18.078       &6.1334e-01     &7.3343e-02     &0.15634        &0.01870        &0.21243        &V      &8.363  \\
2058196.52720   &205819.65270  &18.626  &+0.152 &-0.133   &18.268       &5.0287e-01     &6.5507e-02     &0.12818        &0.01670        &0.17828        &V      &7.676  \\
2515573.53320   &251557.35330  &19.156  &+0.248 &-0.202   &18.637       &3.0871e-01     &6.3040e-02     &0.07869        &0.01607        &0.12690        &V      &4.897  \\
3074589.87390   &307458.98740  &20.848  &+nan   &-0.801   &19.271       &6.5006e-02     &7.0919e-02     &0.01657        &0.01808        &0.07080        &V      &0.917  \\
3757832.06820   &375783.20680  &20.215  &+1.081 &-0.531   &19.063       &1.1638e-01     &7.3368e-02     &0.02966        &0.01870        &0.08577        &V      &1.586  \\
\hline                                                                                    4009.89170     &400.98920     &19.297  &+0.265 &-0.213   &18.753        &7.4653e-01     &1.6183e-01     &0.06881        &0.01492        &0.11356        &B      &4.613  \\
5990.08510      &599.00850     &19.024  &+0.233 &-0.191   &18.529       &9.5987e-01     &1.8513e-01     &0.08848        &0.01706        &0.13967        &B      &5.185  \\
24405.30400     &2440.53040    &18.855  &+0.159 &-0.138   &18.484       &1.1215e+00     &1.5251e-01     &0.10337        &0.01406        &0.14555        &B      &7.353  \\
29828.70490     &2982.87050    &19.222  &+0.159 &-0.139   &18.850       &8.0004e-01     &1.0909e-01     &0.07374        &0.01006        &0.10391        &B      &7.334  \\
36457.30590     &3645.73060    &18.831  &+0.327 &-0.251   &18.205       &1.1473e+00     &2.9823e-01     &0.10575        &0.02749        &0.18822        &B      &3.847  \\
44558.92950     &4455.89290    &18.961  &+0.369 &-0.275   &18.285       &1.0177e+00     &2.9314e-01     &0.09380        &0.02702        &0.17486        &B      &3.472  \\
54460.91380     &5446.09140    &18.917  &+0.248 &-0.202   &18.398       &1.0600e+00     &2.1629e-01     &0.09771        &0.01994        &0.15752        &B      &4.901  \\
66563.33910     &6656.33390    &19.073  &+0.241 &-0.197   &18.565       &9.1803e-01     &1.8253e-01     &0.08462        &0.01682        &0.13509        &B      &5.029  \\
81355.19220     &8135.51920    &18.826  &+0.301 &-0.235   &18.233       &1.1523e+00     &2.7905e-01     &0.10621        &0.02572        &0.18337        &B      &4.129  \\
99434.12380     &9943.41240    &19.397  &+0.383 &-0.283   &18.704       &6.8127e-01     &2.0271e-01     &0.06280        &0.01868        &0.11885        &B      &3.361  \\
121530.59580    &12153.05960   &19.719  &+0.487 &-0.335   &18.921       &5.0634e-01     &1.8312e-01     &0.04667        &0.01688        &0.09731        &B      &2.765  \\
148537.39480    &14853.73950   &nan     &+nan   &-nan     &19.610       &-1.504e-01     &2.3678e-01     &-0.0139        &0.02183        &0.05161        &B      &-0.635 \\
181545.70480    &18154.57050   &20.157  &+1.017 &-0.516   &19.030       &3.3826e-01     &2.0569e-01     &0.03118        &0.01896        &0.08806        &B      &1.644  \\
221889.19480    &22188.91950   &20.025  &+0.378 &-0.280   &19.339       &3.8196e-01     &1.1222e-01     &0.03521        &0.01034        &0.06624        &B      &3.404  \\
271197.90470    &27119.79050   &19.895  &+0.378 &-0.280   &19.209       &4.3049e-01     &1.2652e-01     &0.03968        &0.01166        &0.07467        &B      &3.403  \\
331464.10580    &33146.41060   &19.900  &+0.355 &-0.267   &19.240       &4.2838e-01     &1.1949e-01     &0.03949        &0.01101        &0.07253        &B      &3.585  \\
405122.79590    &40512.27960   &19.681  &+2.190 &-0.678   &18.290       &5.2436e-01     &4.5462e-01     &0.04833        &0.04190        &0.17404        &B      &1.153  \\
605183.43590    &60518.34360   &19.320  &+0.193 &-0.164   &18.887       &7.3107e-01     &1.1930e-01     &0.06739        &0.01100        &0.10037        &B      &6.128  \\
739668.64390    &73966.86440   &18.832  &+0.115 &-0.104   &18.545       &1.1463e+00     &1.1548e-01     &0.10566        &0.01064        &0.13759        &B      &9.926  \\
904039.45360    &90403.94540   &18.686  &+0.095 &-0.088   &18.442       &1.3110e+00     &1.1021e-01     &0.12084        &0.01016        &0.15131        &B      &11.895 \\
1104937.11000   &110493.71100  &18.769  &+0.086 &-0.080   &18.545       &1.2145e+00     &9.2847e-02     &0.11195        &0.00856        &0.13762        &B      &13.081 \\
1350478.69000   &135047.86900  &19.069  &+0.107 &-0.098   &18.799       &9.2124e-01     &8.6730e-02     &0.08491        &0.00799        &0.10890        &B      &10.622 \\
1650585.06550   &165058.50660  &19.482  &+0.181 &-0.155   &19.070       &6.2981e-01     &9.6877e-02     &0.05805        &0.00893        &0.08484        &B      &6.501  \\
2017381.74670   &201738.17470  &19.913  &+0.221 &-0.184   &19.434       &4.2363e-01     &7.8164e-02     &0.03905        &0.00720        &0.06066        &B      &5.420  \\
2465688.80160   &246568.88020  &20.497  &+0.401 &-0.292   &19.785       &2.4728e-01     &7.6390e-02     &0.02279        &0.00704        &0.04392        &B      &3.237  \\
3013619.64630   &301361.96460  &20.597  &+0.551 &-0.364   &19.744       &2.2561e-01     &8.9792e-02     &0.02080        &0.00828        &0.04562        &B      &2.513  \\
3683312.90110   &368331.29010  &21.274  &+1.373 &-0.587   &20.027       &1.2092e-01     &8.6785e-02     &0.01115        &0.00800        &0.03514        &B      &1.393  \\
\hline 
\end{tabular}
\end{sideways}}
\medskip
\medskip
\medskip

\caption{\label{obs-log}  Log of  the \emph{Swift}/UVOT observations
used in this work. Columns 1 and 2 show the mid-time of each observations and the time binning divided by $2$; columns 3 to 6 report the magnitude values in the AB system with positive and negative errors or a 3-$\sigma$ upper limit in case of a non-detection; columns 7 and 8 show the count rates with errors, columns 9-11 the flux in mJy with errors and upper limits (in case of non-detections); columns 12 and display the UVOT filter and the sgnal-to-noise ratio of the measure, respectively.}
\end{table*}

\begin{table*} \centering
\resizebox{\textwidth}{!}{%
\begin{sideways}
\begin{tabular}{|l|l| c|c|c|c|c|c|c|c|c|c|c|}
\hline 
 Tmid (s)               &Terr (s)        &AB Mag        &Mag E+ &Mag E-   &AB MagU ($3\sigma$)     &Rate (cts/s)         &RateErr (cts/s)        &Flux (mJy)           &Flux Err (mJy)      &FluxUL ($3\sigma$)   &Filter      &S/N    \\
\hline                                                                                                      346.74660    &34.67470      &19.120  &+0.272 &-0.217   &18.567       &1.2471e+00     &2.7593e-01     &0.08196        &0.01813        &0.13636        &U      &4.520  \\
5755.91820      &575.59180     &18.746  &+0.109 &-0.099   &18.473       &1.7604e+00     &1.6745e-01     &0.11570        &0.01100        &0.14871        &U      &10.513 \\
15698.76520     &1569.87650    &18.776  &+0.073 &-0.068   &18.583       &1.7131e+00     &1.1106e-01     &0.11258        &0.00730        &0.13448        &U      &15.425 \\
35032.10200     &3503.21020    &18.683  &+0.124 &-0.111   &18.378       &1.8660e+00     &2.0152e-01     &0.12263        &0.01324        &0.16236        &U      &9.260  \\
42817.01360     &4281.70140    &19.174  &+0.247 &-0.201   &18.656       &1.1869e+00     &2.4182e-01     &0.07801        &0.01589        &0.12568        &U      &4.908  \\
52331.90550     &5233.19050    &18.865  &+0.141 &-0.125   &18.527       &1.5774e+00     &1.9190e-01     &0.10367        &0.01261        &0.14150        &U      &8.220  \\
63961.21780     &6396.12180    &19.005  &+0.146 &-0.128   &18.658       &1.3867e+00     &1.7409e-01     &0.09114        &0.01144        &0.12546        &U      &7.965  \\
78174.82170     &7817.48220    &19.818  &+0.300 &-0.235   &19.226       &6.5565e-01     &1.5846e-01     &0.04309        &0.01041        &0.07433        &U      &4.138  \\
95547.00430     &9554.70040    &19.607  &+0.279 &-0.222   &19.043       &7.9649e-01     &1.8073e-01     &0.05234        &0.01188        &0.08798        &U      &4.407  \\
116779.67200    &11677.96720   &19.962  &+0.295 &-0.232   &19.377       &5.7443e-01     &1.3654e-01     &0.03775        &0.00897        &0.06467        &U      &4.207  \\
142730.71020    &14273.07100   &20.637  &+0.713 &-0.427   &19.667       &3.0832e-01     &1.4846e-01     &0.02026        &0.00976        &0.04953        &U      &2.077  \\
213215.01150    &21321.50120   &21.026  &+0.452 &-0.318   &20.262       &2.1553e-01     &7.3424e-02     &0.01416        &0.00483        &0.02864        &U      &2.935  \\
260596.12520    &26059.61250   &20.563  &+0.307 &-0.239   &19.963       &3.3033e-01     &8.1260e-02     &0.02171        &0.00534        &0.03773        &U      &4.065  \\
318506.37520    &31850.63750   &20.709  &+0.408 &-0.296   &19.989       &2.8867e-01     &9.0482e-02     &0.01897        &0.00595        &0.03681        &U      &3.190  \\
389285.56970    &38928.55700   &21.431  &+1.812 &-0.645   &20.091       &1.4844e-01     &1.2046e-01     &0.00976        &0.00792        &0.03350        &U      &1.232  \\
581525.35730    &58152.53570   &21.527  &+1.127 &-0.541   &20.357       &1.3588e-01     &8.7740e-02     &0.00893        &0.00577        &0.02623        &U      &1.549  \\
710753.21440    &71075.32140   &21.198  &+0.481 &-0.332   &20.405       &1.8406e-01     &6.5915e-02     &0.01210        &0.00433        &0.02509        &U      &2.792  \\
868698.37320    &86869.83730   &21.048  &+0.452 &-0.318   &20.284       &2.1116e-01     &7.1963e-02     &0.01388        &0.00473        &0.02807        &U      &2.934  \\
1061742.45610   &106174.24560  &21.130  &+0.375 &-0.278   &20.447       &1.9588e-01     &5.7186e-02     &0.01287        &0.00376        &0.02415        &U      &3.425  \\
1297685.22410   &129768.52240  &21.254  &+0.422 &-0.303   &20.519       &1.7475e-01     &5.6332e-02     &0.01148        &0.00370        &0.02259        &U      &3.102  \\
1586059.71840   &158605.97180  &21.874  &+0.920 &-0.491   &20.790       &9.8677e-02     &5.6375e-02     &0.00648        &0.00370        &0.01760        &U      &1.750  \\
1938517.43360   &193851.74340  &21.832  &+0.782 &-0.450   &20.820       &1.0257e-01     &5.2663e-02     &0.00674        &0.00346        &0.01712        &U      &1.948  \\
2369299.08550   &236929.90860  &21.955  &+0.885 &-0.481   &20.888       &9.1584e-02     &5.1052e-02     &0.00602        &0.00336        &0.01608        &U      &1.794  \\
2895809.99340   &289580.99930  &21.747  &+0.715 &-0.428   &20.775       &1.1100e-01     &5.3564e-02     &0.00729        &0.00352        &0.01786        &U      &2.072  \\
3539323.32530   &353932.33250  &22.329  &+2.665 &-0.705   &20.896       &6.4951e-02     &5.9371e-02     &0.00427        &0.00390        &0.01597        &U      &1.094  \\
4325839.61980   &432583.96200  &21.063  &+0.715 &-0.427   &20.092       &2.0826e-01     &1.0046e-01     &0.01369        &0.00660        &0.03349        &U      &2.073  \\
\hline                                                                                                                                                                 
5455.54780      &545.55480     &18.739  &+0.108 &-0.098   &18.469       &1.2140e+00     &1.1468e-01     &0.11570        &0.01093        &0.14849        &UVW1   &10.586 \\
9960.67770      &996.06780     &18.654  &+0.107 &-0.097   &18.385       &1.3134e+00     &1.2308e-01     &0.12517        &0.01173        &0.16036        &UVW1   &10.671 \\
14879.53090     &1487.95310    &18.819  &+0.072 &-0.067   &18.628       &1.1286e+00     &7.2313e-02     &0.10756        &0.00689        &0.12823        &UVW1   &15.607 \\
33203.96470     &3320.39650    &19.109  &+0.196 &-0.166   &18.672       &8.6349e-01     &1.4262e-01     &0.08229        &0.01359        &0.12307        &UVW1   &6.055  \\
40582.62350     &4058.26230    &19.830  &+0.209 &-0.175   &19.371       &4.4450e-01     &7.7929e-02     &0.04236        &0.00743        &0.06464        &UVW1   &5.704  \\
49600.98420     &4960.09840    &19.617  &+0.270 &-0.216   &19.067       &5.4100e-01     &1.1908e-01     &0.05156        &0.01135        &0.08560        &UVW1   &4.543  \\
60623.42520     &6062.34250    &20.376  &+0.273 &-0.218   &19.821       &2.6890e-01     &5.9811e-02     &0.02563        &0.00570        &0.04273        &UVW1   &4.496  \\
74095.29740     &7409.52970    &20.658  &+0.426 &-0.305   &19.920       &2.0743e-01     &6.7320e-02     &0.01977        &0.00642        &0.03902        &UVW1   &3.081  \\
90560.91910     &9056.09190    &20.841  &+0.545 &-0.361   &19.993       &1.7523e-01     &6.9124e-02     &0.01670        &0.00659        &0.03646        &UVW1   &2.535  \\
110685.56780    &11068.55680   &22.052  &+2.300 &-0.685   &20.650       &5.7417e-02     &5.0511e-02     &0.00547        &0.00481        &0.01991        &UVW1   &1.137  \\
135282.36060    &13528.23610   &21.137  &+0.598 &-0.383   &20.247       &1.3345e-01     &5.6500e-02     &0.01272        &0.00538        &0.02887        &UVW1   &2.362  \\
202088.46460    &20208.84650   &21.826  &+0.467 &-0.325   &21.048       &7.0713e-02     &2.4711e-02     &0.00674        &0.00236        &0.01380        &UVW1   &2.862  \\
246997.01230    &24699.70120   &21.966  &+0.481 &-0.332   &21.174       &6.2154e-02     &2.2246e-02     &0.00592        &0.00212        &0.01228        &UVW1   &2.794  \\
301885.23730    &30188.52370   &nan     &+nan   &-nan     &21.395       &-1.5223e-02    &4.0127e-02     &-0.0015        &0.00382        &0.01002        &UVW1   &-0.379 \\
368970.84560    &36897.08460   &22.406  &+0.999 &-0.511   &21.286       &4.1455e-02     &2.4935e-02     &0.00395        &0.00238        &0.01108        &UVW1   &1.663  \\
551178.67050    &55117.86710   &22.547  &+2.477 &-0.696   &21.128       &3.6414e-02     &3.2696e-02     &0.00347        &0.00312        &0.01282        &UVW1   &1.114  \\
673662.81950    &67366.28200   &23.329  &+nan   &-0.809   &21.741       &1.7724e-02     &1.9600e-02     &0.00169        &0.00187        &0.00729        &UVW1   &0.904  \\
823365.66830    &82336.56680   &23.817  &+nan   &-1.138   &21.775       &1.1306e-02     &2.0934e-02     &0.00108        &0.00200        &0.00706        &UVW1   &0.540  \\
1006335.81680   &100633.58170  &22.602  &+0.754 &-0.441   &21.606       &3.4614e-02     &1.7336e-02     &0.00330        &0.00165        &0.00826        &UVW1   &1.997  \\
1229965.99840   &122996.59980  &23.093  &+1.814 &-0.645   &21.753       &2.2029e-02     &1.7884e-02     &0.00210        &0.00170        &0.00721        &UVW1   &1.232  \\
1503291.77580   &150329.17760  &23.625  &+nan   &-0.874   &21.943       &1.3488e-02     &1.6672e-02     &0.00129        &0.00159        &0.00605        &UVW1   &0.809  \\
1837356.61480   &183735.66150  &24.392  &+nan   &-1.338   &22.097       &6.6540e-03     &1.6156e-02     &0.00063        &0.00154        &0.00525        &UVW1   &0.412  \\
2245658.08480   &224565.80850  &24.009  &+nan   &-1.070   &22.058       &9.4680e-03     &1.5889e-02     &0.00090        &0.00151        &0.00545        &UVW1   &0.596  \\
2744693.21470   &274469.32150  &23.700  &+nan   &-0.899   &21.981       &1.2592e-02     &1.6238e-02     &0.00120        &0.00155        &0.00584        &UVW1   &0.775  \\
3354625.04020   &335462.50400  &nan     &+nan   &-nan     &22.180       &-3.7920e-03    &1.8277e-02     &-0.0004        &0.00174        &0.00486        &UVW1   &-0.207 \\
4100097.27140   &410009.72710  &24.259  &+nan   &-1.623   &21.619       &7.5220e-03     &2.6020e-02     &0.00072        &0.00248        &0.00816        &UVW1   &0.289  \\
\hline 
\end{tabular}
\end{sideways}}
\medskip
\medskip
\medskip
\linebreak
- continued
\end{table*}

\begin{table*} \centering
\resizebox{\textwidth}{!}{%
\begin{sideways}
\begin{tabular}{|l|l| c|c|c|c|c|c|c|c|c|c|c|}
\hline 
 Tmid (s)               &Terr (s)        &AB Mag        &Mag E+ &Mag E-   &AB MagU ($3\sigma$)     &Rate (cts/s)         &RateErr (cts/s)        &Flux (mJy)           &Flux Err (mJy)      &FluxUL ($3\sigma$)   &Filter      &S/N    \\
\hline           
5227.53450      &522.75350     &18.871  &+0.132 &-0.117   &18.551       &7.3714e-01     &8.4117e-02     &0.10296        &0.01175        &0.13821        &UVM2   &8.763  \\
9544.37370      &954.43740     &18.624  &+0.074 &-0.069   &18.429       &9.2513e-01     &6.0966e-02     &0.12922        &0.00852        &0.15477        &UVM2   &15.174 \\
38886.48290     &3888.64830    &20.174  &+0.323 &-0.249   &19.553       &2.2194e-01     &5.7087e-02     &0.03100        &0.00797        &0.05492        &UVM2   &3.888  \\
47527.92360     &4752.79240    &20.680  &+0.324 &-0.249   &20.057       &1.3929e-01     &3.5970e-02     &0.01946        &0.00502        &0.03453        &UVM2   &3.872  \\
58089.68430     &5808.96840    &20.817  &+0.616 &-0.390   &19.914       &1.2276e-01     &5.3130e-02     &0.01715        &0.00742        &0.03941        &UVM2   &2.311  \\
70998.50310     &7099.85030    &21.040  &+0.490 &-0.337   &20.239       &1.0001e-01     &3.6342e-02     &0.01397        &0.00508        &0.02920        &UVM2   &2.752  \\
86775.94820     &8677.59480    &21.624  &+0.697 &-0.421   &20.664       &5.8392e-02     &2.7658e-02     &0.00816        &0.00386        &0.01975        &UVM2   &2.111  \\
106059.49230    &10605.94920   &21.276  &+0.458 &-0.321   &20.506       &8.0470e-02     &2.7706e-02     &0.01124        &0.00387        &0.02285        &UVM2   &2.904  \\
129628.26830    &12962.82680   &22.151  &+0.954 &-0.500   &21.051       &3.5958e-02     &2.1017e-02     &0.00502        &0.00294        &0.01383        &UVM2   &1.711  \\
193642.22800    &19364.22280   &23.031  &+1.506 &-0.608   &21.751       &1.5983e-02     &1.1990e-02     &0.00223        &0.00167        &0.00726        &UVM2   &1.333  \\
236673.83420    &23667.38340   &24.237  &+nan   &-1.151   &22.179       &5.2632e-03     &9.9234e-03     &0.00074        &0.00139        &0.00489        &UVM2   &0.530  \\
289268.01960    &28926.80200   &22.215  &+0.802 &-0.456   &21.192       &3.3871e-02     &1.7682e-02     &0.00473        &0.00247        &0.01214        &UVM2   &1.916  \\
353549.80180    &35354.98020   &23.198  &+1.510 &-0.608   &21.917       &1.3703e-02     &1.0291e-02     &0.00191        &0.00144        &0.00623        &UVM2   &1.332  \\
645507.25120    &64550.72510   &26.843  &+nan   &-3.284   &22.402       &4.7730e-04     &9.3522e-03     &0.00007        &0.00131        &0.00399        &UVM2   &0.051  \\
788953.30700    &78895.33070   &nan     &+nan   &-nan     &22.367       &-1.6660e-04    &9.8760e-03     &-0.0000        &0.00138        &0.00412        &UVM2   &-0.017 \\
964276.26420    &96427.62640   &nan     &+nan   &-nan     &22.970       &-8.1047e-03    &8.3352e-03     &-0.0011        &0.00116        &0.00236        &UVM2   &-0.972 \\
1178559.87840   &117855.98780  &nan     &+nan   &-nan     &22.704       &-3.5571e-03    &8.3856e-03     &-0.0005        &0.00117        &0.00302        &UVM2   &-0.424 \\
1440462.07360   &144046.20740  &nan     &+nan   &-nan     &22.730       &-2.7624e-03    &7.9510e-03     &-0.0004        &0.00111        &0.00295        &UVM2   &-0.347 \\
1760564.75670   &176056.47570  &nan     &+nan   &-nan     &22.937       &-6.4584e-03    &7.9613e-03     &-0.0009        &0.00111        &0.00243        &UVM2   &-0.811 \\
2151801.36930   &215180.13690  &nan     &+nan   &-nan     &22.691       &-1.0908e-03    &7.6468e-03     &-0.0002        &0.00107        &0.00305        &UVM2   &-0.143 \\
2629979.45130   &262997.94510  &nan     &+nan   &-nan     &22.875       &-4.2596e-03    &7.5681e-03     &-0.0006        &0.00106        &0.00258        &UVM2   &-0.563 \\
3214419.32940   &321441.93290  &nan     &+nan   &-nan     &23.263       &-1.1364e-02    &8.0921e-03     &-0.0016        &0.00113        &0.00180        &UVM2   &-1.404 \\
3928734.73590   &392873.47360  &nan     &+nan   &-nan     &23.350       &-1.5664e-02    &9.1911e-03     &-0.0022        &0.00128        &0.00166        &UVM2   &-1.704 \\
\hline                                                                                                                                                                  
4772.62190      &477.26220     &18.679  &+0.108 &-0.099   &18.407       &1.4872e+00     &1.4124e-01     &0.12241        &0.01163        &0.15729        &UVW2   &10.530 \\
5833.20460      &583.32050     &18.685  &+0.109 &-0.099   &18.410       &1.4793e+00     &1.4181e-01     &0.12176        &0.01167        &0.15678        &UVW2   &10.431 \\
19445.01410     &1944.50140    &19.341  &+0.094 &-0.087   &19.100       &8.0815e-01     &6.7178e-02     &0.06652        &0.00553        &0.08311        &UVW2   &12.030 \\
29047.49010     &2904.74900    &20.249  &+0.974 &-0.505   &19.140       &3.5029e-01     &2.0746e-01     &0.02883        &0.01708        &0.08006        &UVW2   &1.688  \\
35502.48790     &3550.24880    &21.276  &+nan   &-0.945   &19.494       &1.3602e-01     &1.8869e-01     &0.01120        &0.01553        &0.05779        &UVW2   &0.721  \\
43391.92970     &4339.19300    &21.090  &+0.365 &-0.273   &20.418       &1.6149e-01     &4.6084e-02     &0.01329        &0.00379        &0.02467        &UVW2   &3.504  \\
53034.58070     &5303.45810    &21.016  &+0.353 &-0.266   &20.359       &1.7279e-01     &4.7935e-02     &0.01422        &0.00395        &0.02606        &UVW2   &3.605  \\
64820.04310     &6482.00430    &20.881  &+0.349 &-0.264   &20.228       &1.9564e-01     &5.3775e-02     &0.01610        &0.00443        &0.02938        &UVW2   &3.638  \\
79224.49710     &7922.44970    &21.475  &+0.319 &-0.246   &20.860       &1.1321e-01     &2.8794e-02     &0.00932        &0.00237        &0.01643        &UVW2   &3.932  \\
96829.94100     &9682.99410    &22.308  &+0.815 &-0.460   &21.278       &5.2556e-02     &2.7751e-02     &0.00433        &0.00228        &0.01118        &UVW2   &1.894  \\
118347.70560    &11834.77060   &21.827  &+0.328 &-0.251   &21.201       &8.1848e-02     &2.1317e-02     &0.00674        &0.00175        &0.01200        &UVW2   &3.840  \\
144647.19570    &14464.71960   &22.613  &+1.164 &-0.549   &21.430       &3.9694e-02     &2.6110e-02     &0.00327        &0.00215        &0.00971        &UVW2   &1.520  \\
216077.90970    &21607.79100   &22.487  &+0.488 &-0.336   &21.689       &4.4575e-02     &1.6143e-02     &0.00367        &0.00133        &0.00766        &UVW2   &2.761  \\
264095.22300    &26409.52230   &22.321  &+0.439 &-0.312   &21.570       &5.1938e-02     &1.7287e-02     &0.00428        &0.00142        &0.00854        &UVW2   &3.004  \\
322783.05030    &32278.30500   &22.181  &+0.417 &-0.301   &21.452       &5.9123e-02     &1.8862e-02     &0.00487        &0.00155        &0.00952        &UVW2   &3.135  \\
394512.61700    &39451.26170   &22.520  &+1.362 &-0.586   &21.276       &4.3258e-02     &3.0922e-02     &0.00356        &0.00255        &0.01120        &UVW2   &1.399  \\
589333.66250    &58933.36620   &24.910  &+nan   &-1.572   &22.329       &4.7880e-03     &1.5589e-02     &0.00039        &0.00128        &0.00424        &UVW2   &0.307  \\
720296.69860    &72029.66990   &23.776  &+nan   &-0.811   &22.185       &1.3602e-02     &1.5102e-02     &0.00112        &0.00124        &0.00485        &UVW2   &0.901  \\
880362.63160    &88036.26320   &23.556  &+2.878 &-0.714   &22.110       &1.6650e-02     &1.5474e-02     &0.00137        &0.00127        &0.00519        &UVW2   &1.076  \\
1075998.77190   &107599.87720  &23.587  &+1.626 &-0.624   &22.282       &1.6181e-02     &1.2562e-02     &0.00133        &0.00103        &0.00443        &UVW2   &1.288  \\
1315109.61020   &131510.96100  &23.474  &+1.240 &-0.564   &22.265       &1.7972e-02     &1.2235e-02     &0.00148        &0.00101        &0.00450        &UVW2   &1.469  \\
1607356.19020   &160735.61900  &24.002  &+nan   &-0.841   &22.368       &1.1042e-02     &1.2909e-02     &0.00091        &0.00106        &0.00410        &UVW2   &0.855  \\
1964546.45470   &196454.64550  &25.797  &+nan   &-2.040   &22.680       &2.1150e-03     &1.1736e-02     &0.00017        &0.00097        &0.00307        &UVW2   &0.180  \\
2401112.33350   &240111.23330  &23.777  &+2.001 &-0.663   &22.410       &1.3584e-02     &1.1433e-02     &0.00112        &0.00094        &0.00394        &UVW2   &1.188  \\
2934692.85200   &293469.28520  &24.076  &+nan   &-0.859   &22.415       &1.0322e-02     &1.2446e-02     &0.00085        &0.00102        &0.00392        &UVW2   &0.829  \\
3586846.81920   &358684.68190  &nan     &+nan   &-nan     &22.850       &-6.5630e-03    &1.2830e-02     &-0.0005        &0.00106        &0.00263        &UVW2   &-0.512 \\
4383923.89010   &438392.38900  &23.881  &+nan   &-1.222   &21.732       &1.2344e-02     &2.5684e-02     &0.00102        &0.00211        &0.00736        &UVW2   &0.481  \\
\hline 
\end{tabular}
\end{sideways}}
\medskip
\medskip
\medskip
\linebreak
- continued
\end{table*}

\end{document}